\pdfoutput=1 % only if pdf/png/jpg images are used
\documentclass[cits]{JINST}
\usepackage{amsmath}
\newtheorem{theorem}{Theorem}

\title{MiX: A Position Sensitive Dual-Phase Liquid Xenon Detector}

\author{S. Stephenson$^a$, J. Haefner$^a$, Q. Lin$^b$, K. Ni$^b$, K. Pushkin$^{a,b}$, R.~Raymond$^a$, M. Schubnell$^a$, N. Shutty$^a$, G. Tarl\'{e}$^a$, C. Weaverdyck$^a$, and W. Lorenzon$^a$\thanks{Corresponding author.} \\
\llap{$^a$}Randall Laboratory of Physics, University of Michigan, \\
Ann Arbor, Michigan 48109-1040, U.S.A.\\
\llap{$^b$}INPAC, Department of Physics \& Astronomy and Shanghai Key Laboratory for Particle Physics and Cosmology, Shanghai Jiao Tong University,\\
800 Dongchuan Road, Shanghai, 200240, P.R. China\\

E-mail: \email{lorenzon@umich.edu}\\
}

\abstract{The need for precise characterization of dual-phase xenon detectors has grown as
the technology has matured into a state of high efficacy for rare event searches. The Michigan Xenon detector was constructed to study the microphysics of particle interactions in liquid xenon across a large energy range in an effort to probe aspects of radiation detection in liquid xenon. We report the design and performance of a small 3D position sensitive dual-phase liquid xenon time projection chamber with high light yield ($L_y^{122}=15.2\,$pe/keV at zero field), long electron lifetime ($\tau > 200\,\mu$s), and excellent energy resolution ($\sigma/E = 1\%$ for 1,333\,keV gamma rays in a drift field of 200\,V/cm). Liquid xenon time projection chambers with such high energy resolution may find applications not only in dark matter direct detection searches, but also in neutrinoless double beta decay experiments and other applications.}

\keywords{Dark Matter detectors; Time Projection Chambers; Noble liquid dual phase detectors, Liquid xenon target, Gamma spectroscopy}

\begin{document}
\section{Introduction}\label{sec:intro}

Dual-phase liquid xenon time projection chambers (LXeTPCs) are versatile radiation detection devices capable of both calorimetry and imaging of particle interactions within a condensed xenon target. Particularly fruitful applications have been found in dark matter searches~\cite{XENON100-08,ZEPLINIII,LUX,PandaX}, neutrinoless double beta decay detectors~\cite{EXO}, gamma-ray physics experiments~\cite{LXeGRIT}, medical imaging~\cite{med-image} and recently in neutron detection for Homeland Security applications~\cite{HLS}.

While detection of particles in dark matter searches requires experiments to use ever increasing xenon target volumes (i.e.\ multi-ton scale) and to retreat deep underground where background rates are low, extensive studies of xenon microphysics and LXeTPC detector characteristics are achievable with smaller apparatus on the Earth's surface. Small LXeTPCs that offer high light yield, 3D position sensitivity, and operational flexibility make an excellent platform to carry out LXe microphysics studies which probe the effects of quanta production in LXeTPCs, such as signal quenching, recombination and anti-correlation, extraction efficiency, energy resolution, light yield, charge yield, electric vs.\ nuclear recoil (ER/NR) discrimination, and more. It is most useful to study these effects as a function of deposited energy, interaction type, and with varied TPC electric fields.

LXeTPCs are able to achieve precise interaction vertex reconstruction and good calorimetry due to dual signal readout capabilities. During particle interactions in a LXe target, both ionized and excited xenon atoms are created. Some of these react with surrounding xenon atoms to form short-lived excited dimers (excimers)~\cite{eximers}, while some ionization electrons recombine with ionized xenon atoms, thus quenching a fraction of the charge signal. This leaves an interaction vertex with three populations of quanta: direct excimers (from excited xenon atoms), recombination excimers (from ionized xenon atoms), and free electrons. The direct and recombination excimers decay rapidly~$\mathcal{O}$({10\,ns}) to produce 178\,nm photons, called primary scintillation light (S1). An electric field is applied across the LXe target to drift ionization electrons away from the interaction site, past a gate grid to the liquid-gas interface, where they are extracted in a strong electric field and produce electroluminescence in the gaseous xenon~\cite{Dolg1970}. The scintillation light produced in this process is proportional to the number of ionized electrons and is referred to as S2~\cite{Aprile-S2}. A recent review of liquid xenon particle detectors gives a detailed account of the detection process \cite{Chepel:2012sj}.

Detailed study of liquid xenon in radiation interactions is necessary to better understand the capabilities of LXe as a target in ionization and scintillation detectors. Previous LXeTPCs have been constructed with great success, but more characterization is needed to fully understand the nature of quanta detection in LXe and explore the practical limits in energy and position resolution. The MiX detector was constructed to precisely probe signal production in LXeTPCs, further investigating LXe target properties.

In this paper, the basic features of the MiX apparatus are described in section~\ref{sec:Apparatus}, followed by a discussion of the light yield and energy resolution for gamma rays over a large energy range in the MiX system in section~\ref{sec:results}. Finally, in appendix~\ref{sec:theorem} we present a theorem regarding reconstruction singularities in detectors with symmetric sensor layout that tends to limit the performance of small LXeTPCs with few signal transducers.

\begin{figure}[tbp]
\centering
\includegraphics[width=.75\textwidth]{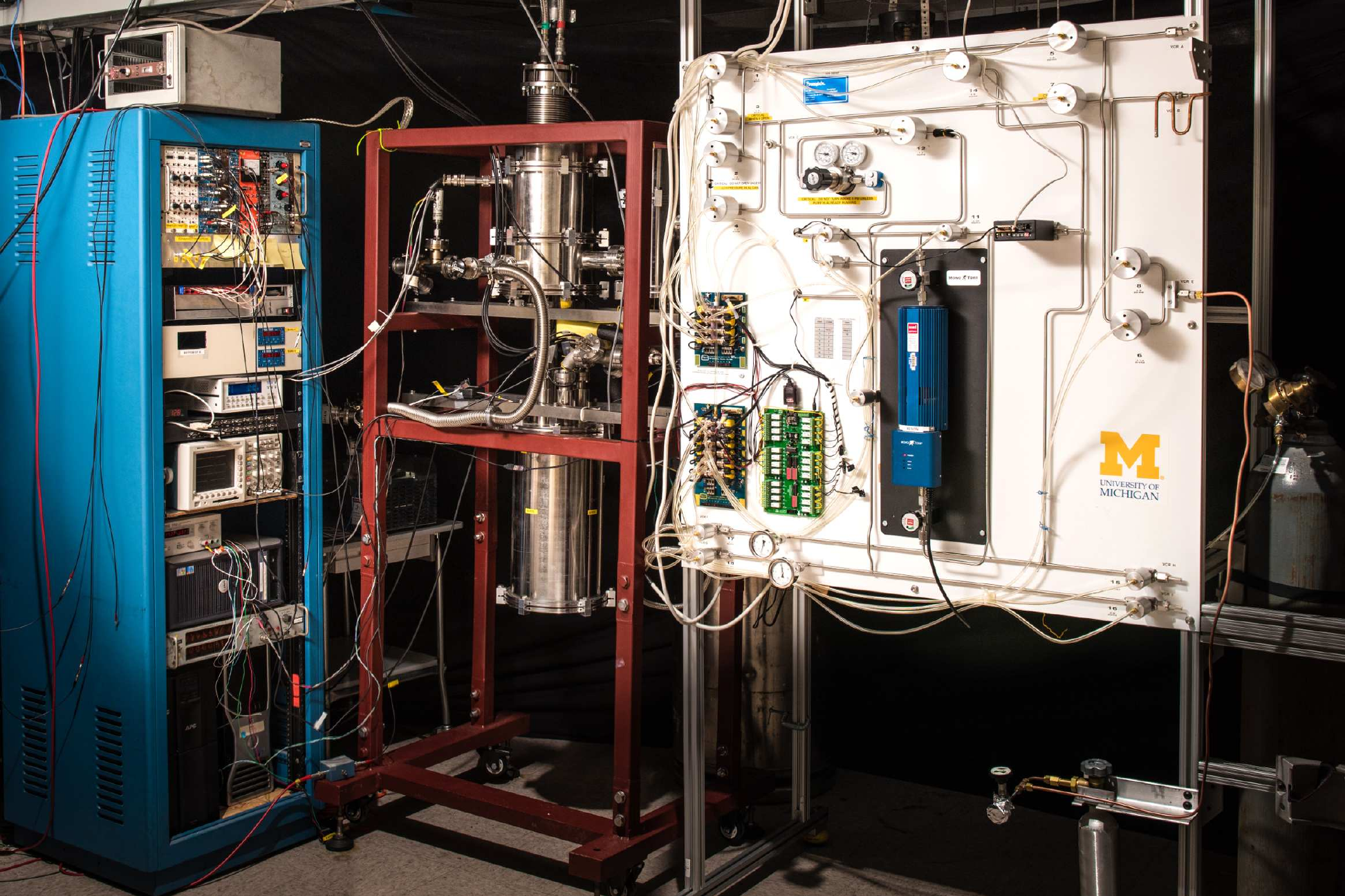}
\caption{Photograph of MiX experimental apparatus. At left, the (blue) rack equipment monitors, controls, and reads out the necessary components in the rest of the system. At center (red frame), the cryostat houses the MiX detector. At right, the (white) gas panel provides storage, handling, and purification of the xenon gas. Details of operation reside in the text.}
\label{fig:System}
\end{figure}

\section{Apparatus}
\label{sec:Apparatus}
The importance of maintaining a stable operational environment for the MiX detector was considered from the beginning as an integral part of the design. It was deemed necessary to provide and maintain not only adequate pressure and temperature control within the cryostat, but also to maintain stable operation of the xenon purification system, the LXe level, the high voltage feed-throughs, the electronics components and the physical space. A photograph of the system is shown in Fig.~\ref{fig:System}.

\paragraph{Cryogenics}

The MiX system consists of a cylindrical inner vessel (6-inch diameter, 12-inch high) that is surrounded with about 10 layers of superinsulation and placed inside a stainless steel vacuum jacket that is evacuated to $10^{-5}$ Torr to reduce heat load on the chilled inner volume. The cooling power is supplied by a 24\,W Joule-Thomson (JT) cooler, model ARS MRCM-150-1. The JT cooler is coupled to the inner volume by an indium-sealed copper condensing head that penetrates into the gaseous xenon chamber. It pulls heat from the condensing head at a constant rate while a Lakeshore temperature control unit (model 340) maintains a constant temperature of the condensation surface
using a PT100 platinum resistor temperature sensor and two resistive heating elements placed in the condensing head.

During filling, gaseous xenon (GXe) stored in high-pressure aluminum bottles passes through a pressure regulator, condenses on the copper condensing head and falls into a stainless steel funnel. The LXe then feeds down to the inner vessel where it cools the detector region.  An emergency LN$_2$ coil is soldered on the copper condensing head. The coil is permanently attached to a LN$_2$ dewar which can supply cooling power in the event of a cooling failure, and adds cooling power during initial cool down.

\paragraph{Gas system}

A gas panel, shown in Fig.~\ref{fig:System}, was assembled to handle the 14\,kg of xenon during storage and normal operation. Two aluminum bottles are hung from load cells for storage of the xenon in gaseous form. One bottle houses all of the xenon while the other acts as a relief/buffer volume in case the pressure elevates too high in the inner vessel. Eighteen Swagelok low and high pressure pneumatic valves are activated by an electronic relay board. The relay board is controlled through a serial-USB interface with a slow control computer to allow remote computer control. The valves default to a normally closed configuration in the event that USB connection is reset or the power cycled.

Liquid xenon is drawn from the detector chamber into the heat exchanger~\cite{Giboni-11} where it evaporates. An MX-808ST-S diaphragm pump pushes the GXe through a hot zirconium SAES PS3-MT3-R-1 getter. The mass flow rate is controlled by a Hastings HFC-D-302 flow controller. The GXe enters the cryostat again, is cooled in the heat exchanger (93\% efficiency~\cite{Giboni-11}), flows into the inner vessel condensation region and is liquified again.

\paragraph{Feed-throughs}

The ability to communicate with and reliably control the outer vacuum and inner xenon space via both optical and electrical means is crucial for operating the TPC. Custom and standard feed-throughs have been used to transport electrical and optical signals between the cryostat interior and the external systems. Two standard MPF ten pin feed-throughs with 1.33'' CF flanges carry low voltage in and out of the detector; one for the cryogenic head located in the outer vacuum and another for the inner vessel temperature sensor, the level meter, and the TPC anode. Two standard four-connector SHV to vacuum BNC (i.e.\ non-standard BNC) feed-throughs are used to transport the PMT high voltage and signals between the inner vessel and the outside world. The SHV-BNC and inner vessel ten pin feed-throughs are mounted to a CF cross and elbow to allow for all PMT, temperature, and level sensor cabling to enter the inner vessel through a single 2.75'' CF flange.

The detector feed-throughs supply negative high voltage $\mathcal{O}$({10\,kV}) to the TPC gate and cathode grids via customized commercial ceramic feed-throughs (interlocking teflon sleeves surround nickel electrodes coupled to 30\,kV Kapton jacketed cables with gold connectors). The PMTs are fed positive high voltage $\mathcal{O}$({1\,kV}) via SHV to vacuum BNC with custom vacuum BNC cable connectors\footnote{It was found that the vacuum side of the standard SHV to vacuum BNC will breakdown at 1\,kV$-3$\,kV in $\mathcal{O}$({1\,atma})  GXe when a standard grounded connector is used. Thus, a machined stainless steel connector with a PTFE core insulator was designed to mate to the unjacketed Kap3 coaxial PMT cables.}. A custom fiber optic feed-through (bare fibre optic core, vacuum epoxy, 1/4'' stainless tubing, Ultratorr fitting) was constructed to cleanly transport optical photons from a pulsed LED into the inner vessel for in situ single photoelectron (pe) calibration of all PMTs.

\paragraph{Slow Control}

A slow control system, developed for use in PandaX~\cite{PandaX-TDR}, was implemented to provide monitoring and control of important system variables. Parameters such as pressures, temperatures, setpoints, LXe liquid level, GXe circulation rate, and valve states are logged asynchronously during twenty second time intervals while the system is running. Plots of system parameters are served to a webpage which can be remotely monitored. An uninterruptible power supply (UPS) maintains power to the system if an outage occurs. If the monitored quantities deviate from a predetermined range, the slow control system sends alarm messages via email and text message within one minute. During emergencies, a liquid nitrogen system is activated to release liquid nitrogen from an LN$_2$ dewar to supply additional cooling power to the condensing head.

\paragraph{Detector}

\begin{figure}[tbp]
\centering
\includegraphics[height=0.5\textwidth]{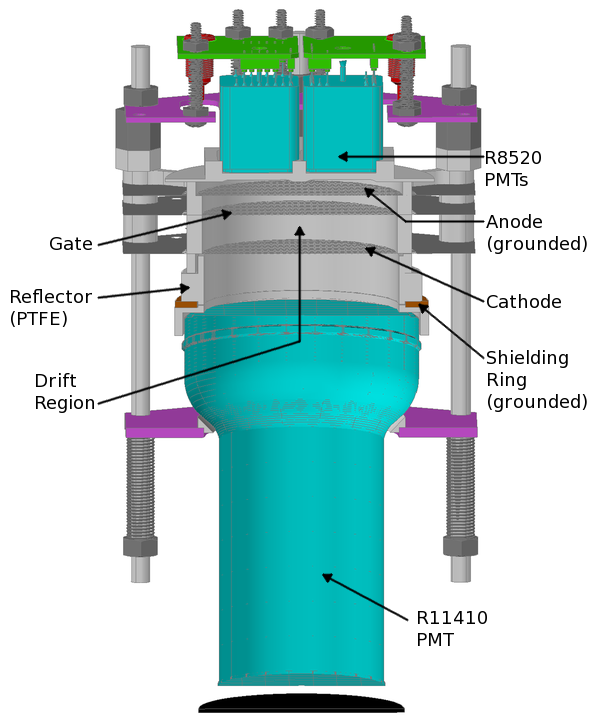} \hspace*{8mm}
\includegraphics[height=0.5\textwidth]{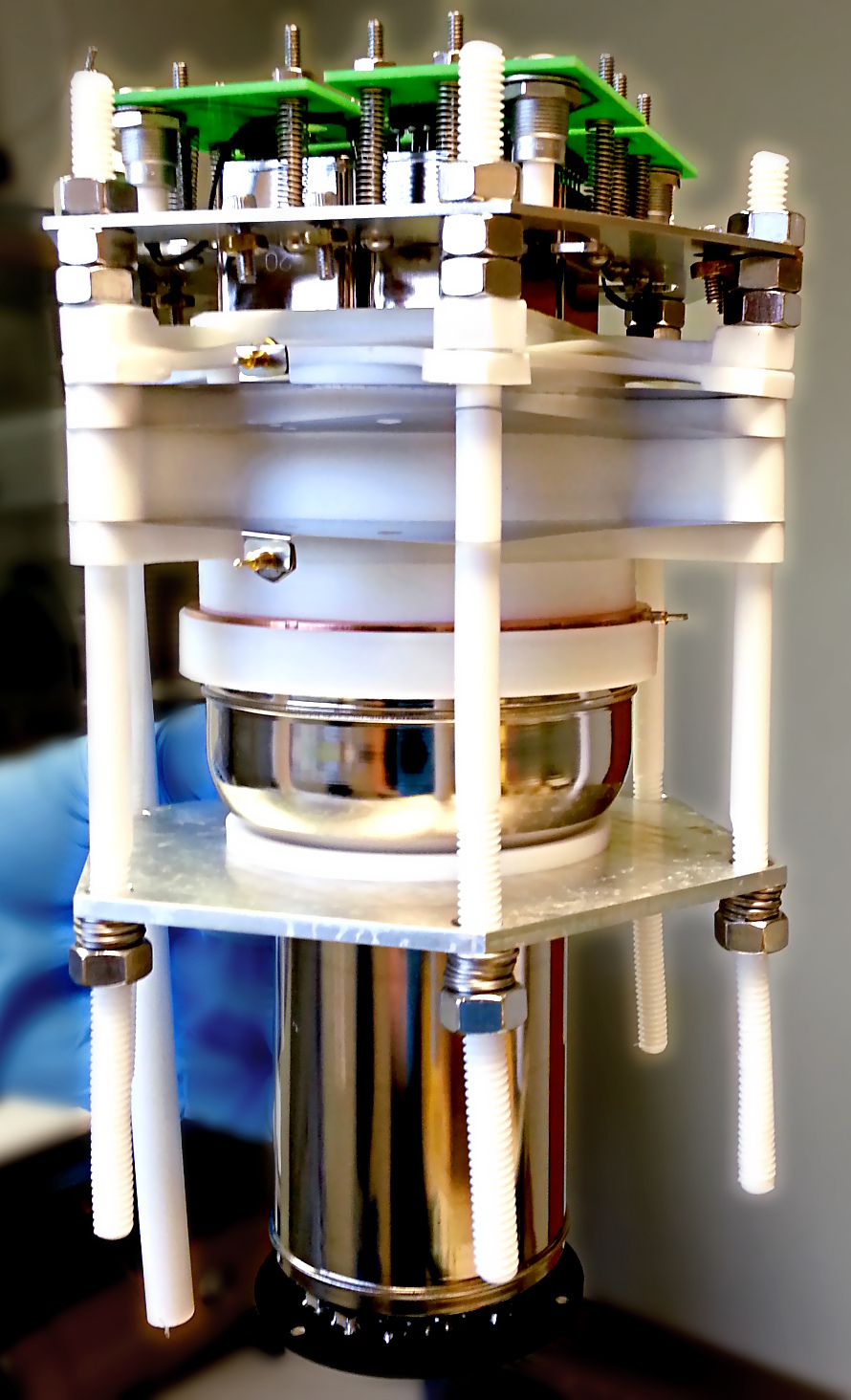}
\caption{Left Panel: Schematic cross section of the MiX TPC. Right Panel: Photograph of the assembled detector. Three stainless steel etched hexagonal meshes define the drift (cathode --  gate: 12\,mm) and the extraction (gate -- anode: 7\,mm) regions. Four Hamamatsu R8520-406 PMTs are used on the top and one Hamamatsu R11410-MOD is used on the bottom. A copper shielding ring protects the bottom PMT.}
\label{fig:TPC}
\end{figure}

\begin{figure}[tbp]
\centering
\includegraphics[width=.55\textwidth]{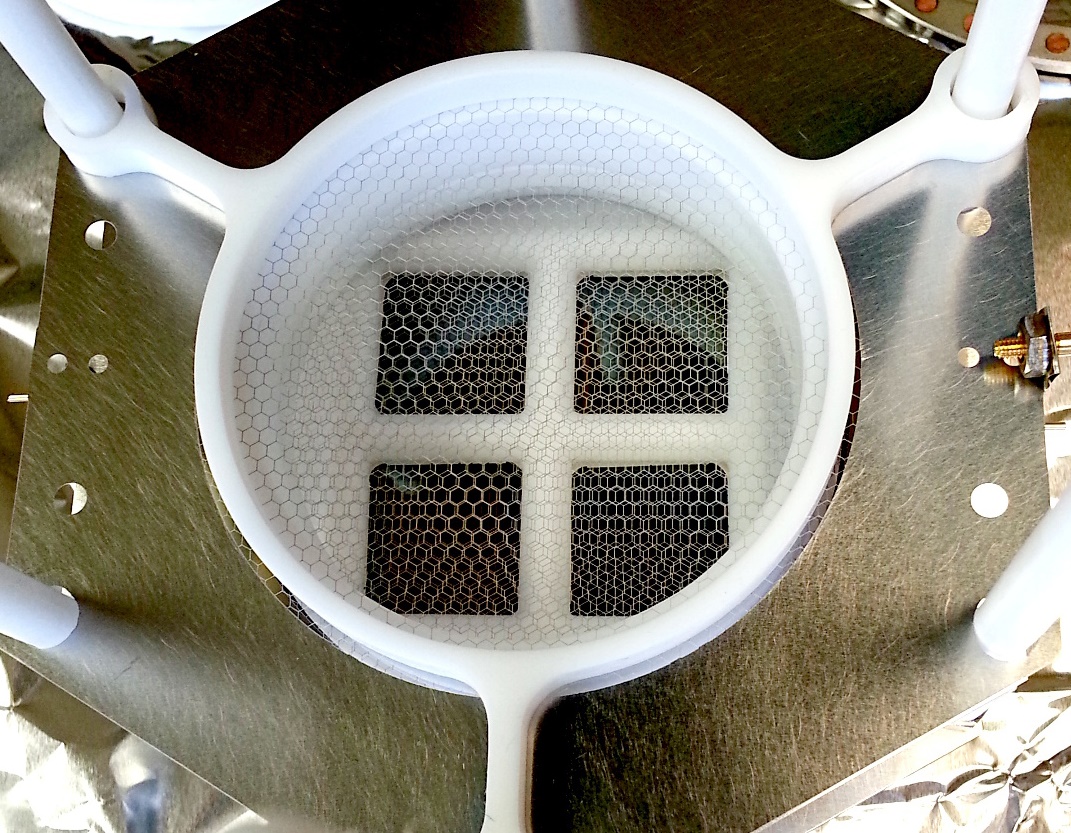}
\caption{Photograph of MiX detector with all three TPC electrodes in place. Each 50\,$\mu$m-thick mesh has a central 76\,mm circular region with chemically etched hexagonal grid structure that provides a 95\% optical transparency. The large unetched region extending beyond the TPC sensitive volume ensures that fringing electric fields induced by the grounded inner vessel surface do not reach the innermost fiducial region.}
\label{fig:Hex-Grids}
\end{figure}

The TPC structure, shown in Fig.~\ref{fig:TPC}, is hung from the top flange of the inner vessel by four stainless steel spring straps. It consists of four top PMTs (TPMTs) and one bottom PMT (BPMT) with high detection efficiency in the VUV region ($>25\%$ at 178\,nm)~\cite{PMTs-12}, three electrode grids (anode, gate, and cathode), one shielding ring for the bottom PMT, and an enveloping virgin PTFE diffuse reflector.  The 133\,mm-diameter hexagonal grids were chemically etched by Great Lakes Engineering from 50\,$\mu$m-thick stainless steel sheet to optimize electric field uniformity. A 2\,mm-pitch hexagonal mesh was formed in the central 76\,mm diameter. The wires were smoothed as a result of the etching, giving a round wire of 50\,$\mu$m diameter and an optical transparency of 95\% (see Fig.~\ref{fig:Hex-Grids}). During operation, the anode grid and shielding ring are connected to ground while the gate and cathode grids are supplied with negative high voltage to provide the appropriate drift and extraction fields.

Liquid xenon fills the detector from the bottom until it partially occupies the space between the gate and anode (the gap region) which contains both gaseous and liquid xenon. A capacitive level meter is used to monitor the height of the liquid-gas interface. The drift region consists of the space between the cathode and gate and is the sensitive volume for radiation detection. Ionization electrons drift past the gate, are extracted from the liquid, and collected on the anode after producing electroluminescence in the gas gap region.  Details of the detector dimensions are listed in Table~\ref{tab:Detector-Parameters}.

\begin{table}[tbhp]
\caption{Detector parameters.}
\label{tab:Detector-Parameters}
\smallskip
\centering
\begin{tabular}{|lc|}
\hline
Detector Parameter & Value \\
\hline
\hline
Diameter & 62.5\,mm \\
TPMT to Anode & 10\,mm \\
Anode to Gate & 7\,mm \\
Gate to Cathode & 12\,mm \\
Cathode to Shield Ring & 17\,mm \\
Shield Ring to BPMT & 3\,mm \\
Grid Pitch & 2\,mm \\
Grid Thickness & $50\,\mu$m \\
Grid Transparency & $95\%$ \\
\hline
\end{tabular}
\end{table}

The top PMT array consists of four Hamamatsu R8520-406 tubes. The 1-inch square, low background, VUV sensitive tubes are arranged in a symmetric four pixel configuration. Each TPMT is affixed to a FR4 positive high voltage base and placed in the top PTFE reflector with only the 20\,mm by 20\,mm active photocathode area exposed to the TPC inner volume. The distribution of the S2 signal in the four TPMTs is used to reconstruct the $XY$ position of an event. The bottom PMT is a 3-inch diameter Hamamatsu R11410-MOD tube. The active photocathode diameter of 64\,mm is matched to the inner TPC diameter (63.5\,mm) to allow for full bottom photocathode coverage of the TPC cylinder. A Cirlex positive HV base is attached to the BPMT and supplied high voltage via Kap3 cabling with custom gold plated HV connections to the BPMT base.

The sensitive volume is designed to be surrounded in near $4\pi$ by either PMT photocathode or virgin PTFE reflector to optimize the light yield. Virgin PTFE has been shown to have above 95\% reflectivity at 178\,nm~\cite{PTFE-refl-04}. The reflector is machined from PTFE sheet and designed to be only as thick as is structurally necessary. The reduction of PTFE mass contained within MiX is done to reduce the low Z material content (in an effort to reduce neutron energy loss when neutrons enter the detector).

The PMT gains were measured from the single photoelectron (SPE) signal induced by an in-situ pulsed green LED system that is coupled to the bare fiber optic cable and pulsed at low amplitude with 10\,ns-wide pulses. The fiber optic cable terminates in the top PTFE reflector, allowing light pulses to diffuse through the PTFE and into the detector volume.  To determine the PMT gains at PMT bias voltages where the SPE signal would be poorly resolved from dark noise, the LED signal is raised to several hundred photoelectrons and held constant while the PMT voltages are lowered incrementally to measure the power law parameters for each PMT. In this way, a low gain setting of $5.0\times10^{5}$ could be selected for each individual TPMT. The BPMT was maintained at a gain of $1.0\times10^{6}$.

\paragraph{Electronics}

Signals induced in the PMTs travel along $50\,\Omega$ SHV cables into custom decouplers. The decoupled signals enter a flash ADC (CAEN V1724) that is mated to an optical
link (CAEN A3818A) in a linux machine. A DRS4 prototype board is used as a digital threshold trigger at 8\,mV offset. Once an event is triggered, a 30\,$\mu$s time interval is captured at a sampling rate of 100\,MS/s for the five PMT channels and stored in a circular buffer. Once the buffer is full, the data are transferred over the optical link and written in binary format (typical event
rates 10\,Hz$-300$\,Hz), using the server internal disk as an initial file buffer. Upon completion, the file is automatically transferred to network storage for backup and offline analysis.

\paragraph{TPC Operation}

\begin{figure}[tbp]
\centering
\includegraphics[width=.8\textwidth]{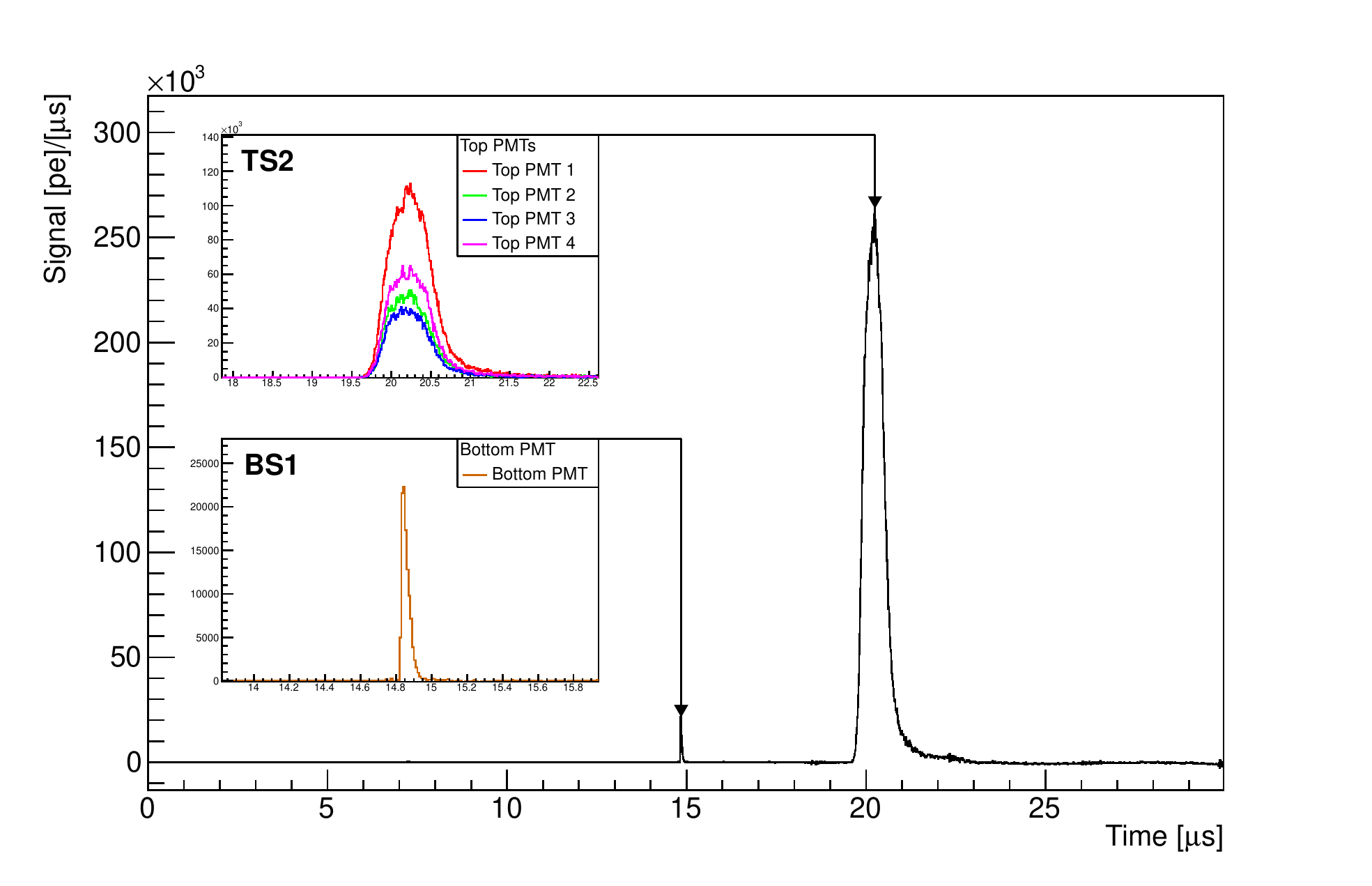}
\caption{A typical event waveform from a 356\,keV gamma ray from $^{133}$Ba. The inserts show the S1 signal for the bottom PMT, and the S2 signal for the 4 individual top PMTs.}
\label{fig:waveform}
\end{figure}

The MiX detector was able to maintain linearity over a large energy range by careful selection of PMT gains and selective use of top and bottom signals. Since the TPMT array sees only a small fraction of the S1 scintillation light ($<15\%$) and has high relative noise (due to the low gain selection driven by S2), the S1 signal is integrated only from the bottom PMT (termed BS1). Additionally, at a gain of $10^{6}$, the BPMT saturates on S2 signals beyond 35\,kpe (corresponding to $\approx25$\,kpe in the summed TPMT array). This means that the BPMT S2 signal is only linear for electromagnetic events with energy below 50\,keV at a 200\,V/cm drift field. Therefore, the S2 signal was chosen to be derived solely from the summed TPMT array (called TS2). This selection allows for a linear response of the detector to better than 5\% in both BS1 and TS2 for electron recoil events of 3\,keV to 1.5\,MeV at a drift field of 200\,V/cm. The TS2 signal will become saturated above 1\,MeV if the drift field is 1\,kV/cm or greater due to the large high-energy TS2 signals resulting from efficient quenching of recombination.

A characteristic high energy signal from the MiX detector is shown in Fig.~\ref{fig:waveform}. Signals are found in the 30\,$\mu$s summed PMT output with a pulse threshold of 0.2\,pe/ns for the results reported here. The BS1 and TS2 signals are typically well formed and easily identified by width using a custom threshold-based pulse finding scheme. Therefore, pulse width-above-threshold is the only criterion applied to determine whether a signal is characterized as an S1 or S2. In particular, a signal is characterized as S1 if the width is above 20\,ns but less than 750\,ns, and as S2 if the width is 750\,ns or above. The selection process can be adjusted to include a width dependence on area if efficiency for proper S1 identification of events with energies beyond 1.5\,MeV is desired. However, determination of the width-dependence-on-energy selection was not studied in depth for this analysis, and the simple width-based selection algorithm is used with great success.

\begin{figure}[tbp]
\centering
\includegraphics[width=0.57\textwidth]{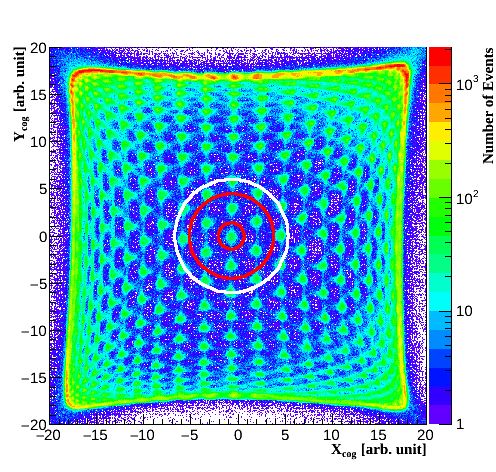}
\caption{Reconstructed $XY$ position based on the center of gravity method for gamma rays from $^{137}$Cs. The dots observed are at a spacing of 2\,mm and correspond to unit cells of the hexagonal anode grid. The existence of the dots allows for a fiducial volume selection with high precision in the $XY$ plane. At large radius, optical warping and position reconstruction degeneracy is observed. The donut-like region (bounded by the two inner, red circles) represents the fiducial region used in the energy resolution analysis, while the white outer ring represents the region used in determining the $\alpha$ and $\beta$ values. See text for more details.}
\label{fig:xy-pos}
\end{figure}

The $XY$ position of each event is reconstructed from the S2 signal distribution on the top PMT array based on a center of gravity algorithm, as shown in Fig.~\ref{fig:xy-pos}. Note that the units of this space are arbitrary, as they represent an optically warped $XY$ plane. The position resolution is good in the center of the detector, as seen from the presence of the pattern imprinted from the hexagonal anode mesh. Near the edges, however, the pattern is warped and reconstruction accuracy is poor and ultimately degenerate (see Appendix~\ref{sec:theorem} for details). The $Z$ position is determined from the time delay between the largest BS1 and largest TS2 signal, since the drift time of electrons in LXe is known as a function of drift field~\cite{Doke-09}.

\begin{figure}[tbp]
\centering
\includegraphics[width=.8\textwidth]{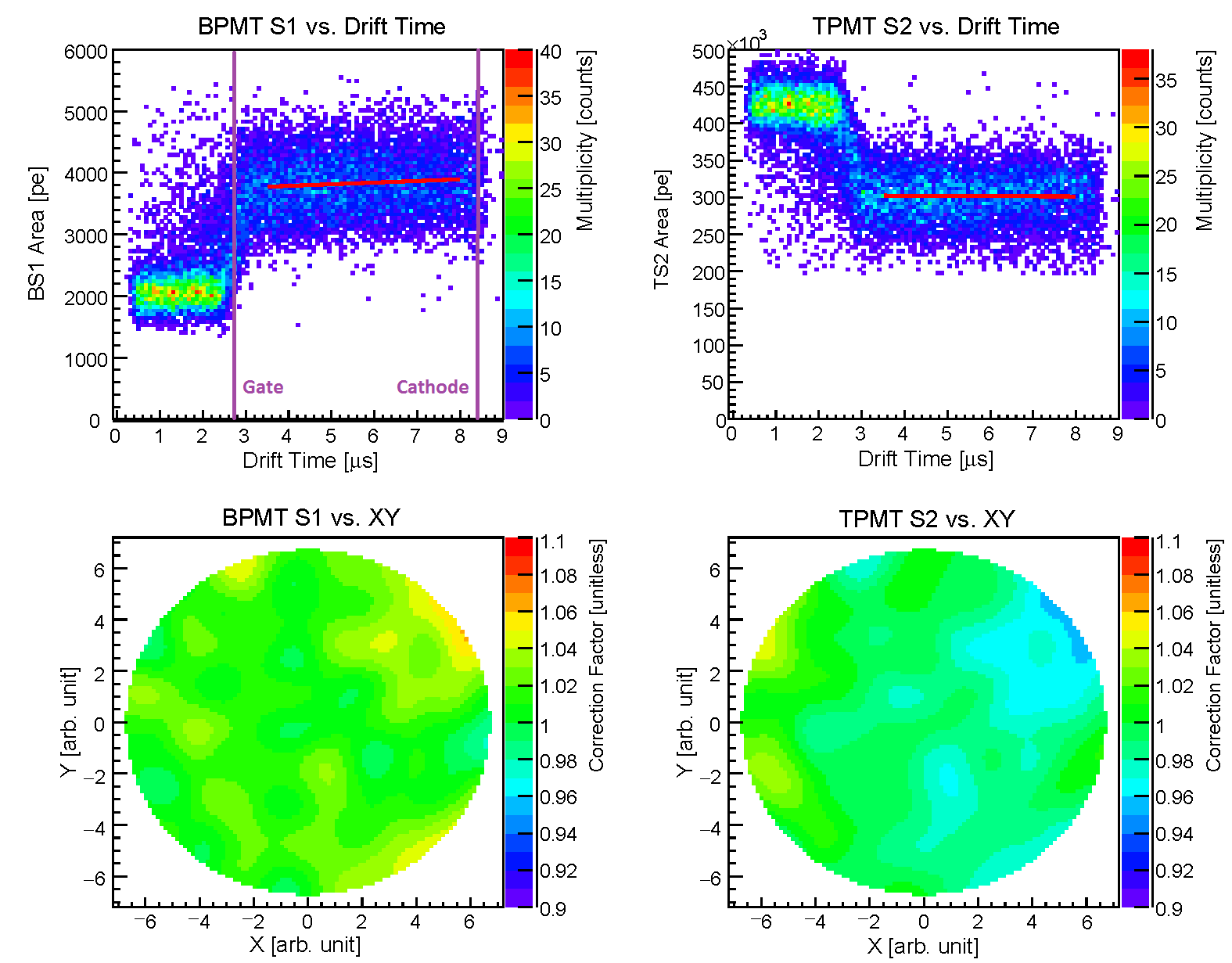}
\caption{Position dependence of signal production in the detector after selection of the 511\,keV photo-absorption peak  of $^{22}${Na}. Top left: BS1 dependence on $Z$ (i.e.\ drift time). The vertical solid lines indicate the positions of the gate and cathode electrodes. Top right: TS2 dependence on $Z$. Bottom left: BS1 dependence on $XY$. Bottom right: TS2 dependence on $XY$. The $XY$ and $Z$ dependencies are factored to produce corrections for both BS1 and TS2 independently.}
\label{fig:Position-dependence}
\end{figure}

Due to non-uniform light collection and light generation inside the sensitive volume, the BS1 and TS2 signals depend on event location. Position-dependent signal correction is achieved by selecting photo-absorption events from a high statistics $^{22}$Na run. Profile plots are generated in $XY$, then smoothed and inverted to produce $XY$ corrections, shown in the bottom panels of Fig.~\ref{fig:Position-dependence}. Additionally, $Z$ profile plots are generated and then fit with linear functions to obtain the $Z$ dependencies, shown in the top panels of Fig.~\ref{fig:Position-dependence}. The two distinct regions in the panels above and below the gate electrode reflect the BS1 and TS2 response to the higher electric field in the extraction and the lower electric field in the drift regions, respectively. In the higher field extraction region recombination is suppressed, leading to a suppression in the primary light signal (BS1), an increase in the number of drift electrons and a corresponding increase in proportional light (TS2). In the lower field drift region, more recombination takes place, leading to an increase in the primary light signal (BS1), a decrease in the number of drift electrons and a corresponding decrease in proportional light (TS2).

Corrections for both BS1 and TS2 are applied as a function of $XY$ and $Z$. The BS1 and TS2 corrections are applied in two steps since many more mono-energetic events would be needed to build a full three dimensional model. The factorization of corrections in this way is not expected to produce a strong systematic effect since the fiducial volume is selected to be small and centrally located. The corrections help the energy resolution of the detector at the sub-percent level. Two fiducial volumes were chosen, as displayed in Fig.~\ref{fig:xy-pos}. The larger-radius circular region was used to determine the anti-correlation axes while the inner symmetric annular region was used to determine energy resolution. The choice for two distinct fiducial regions was based on the the superior energy resolution in the inner region, and the more precise determination of the anti-correlation axes due to the higher statistical sample of events in the larger region. The electron lifetime, extracted from an exponential fit of the TS2 signal (shown in the top right panel in Fig.~\ref{fig:Position-dependence}) was above $200\,\mu$s over the course of the two month data taking period.

\section{Results}
\label{sec:results}

Five electromagnetic radioactive sources were used to produce gamma-ray events inside the MiX sensitive volume in the energy range of 122\,keV to 1.33\,MeV. Furthermore, $^{252}$Cf was used to produce neutron-activated xenon states of $^{131m}$Xe and $^{129m}$Xe at 163.9\,keV and 236.1\,keV, respectively, as listed in Table~\ref{tab:radioactive-sources}. Xenon activation was performed in situ with two $^{252}$Cf sources placed just outside of the outer vessel, taking advantage of the high density of xenon in the liquid state. The combined activity of the $^{252}$Cf sources was 78.7\,$\mu$Ci (April 2015), resulting in a neutron flux of $3.4\times10^5$\,neutrons/s. No data were collected during the week-long xenon activation.

Achieving high energy resolution requires many aspects of the detector to operate predictably. The detector must be extremely stable over periods of data collection which may last from a few hours to a few days. Any change in the electric fields, photomultiplier tube (PMT) gains, or LXe level may modify the conditions for events in the detector and worsen the observed energy resolution. Additionally, a fiducial volume of the detector must be chosen in which the energy resolution is good. Events in the dataset must be classified as either inside or outside of this fiducial volume based on precise $XYZ$ position reconstruction. Finally, small variations in the light collection and signal generation across the detector must be corrected with an $XYZ$ position-dependent signal correction.

\begin{figure}[tbp]
\centering
\includegraphics[width=0.47\textwidth]{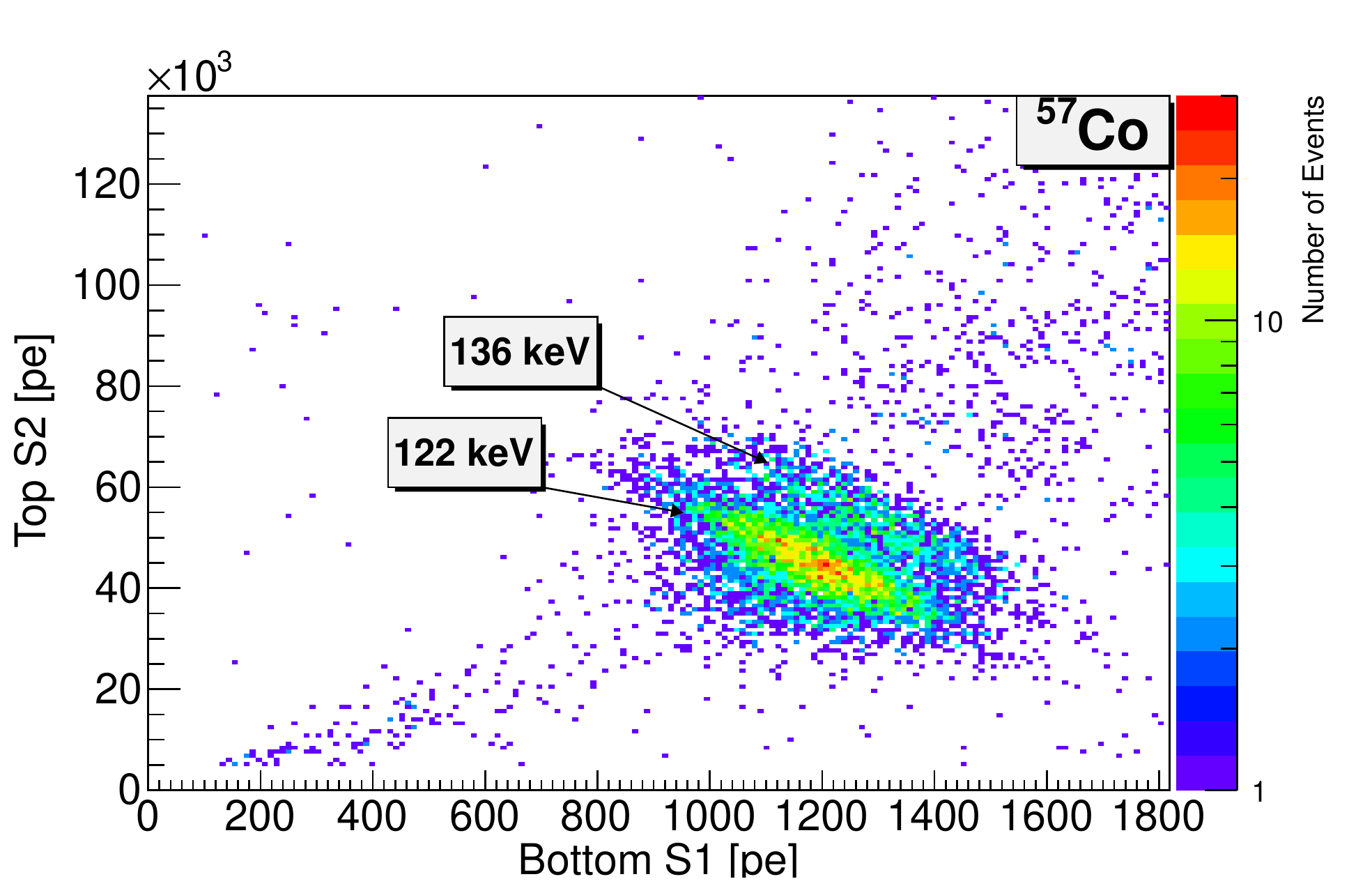} \hspace*{3mm}
\includegraphics[width=0.47\textwidth]{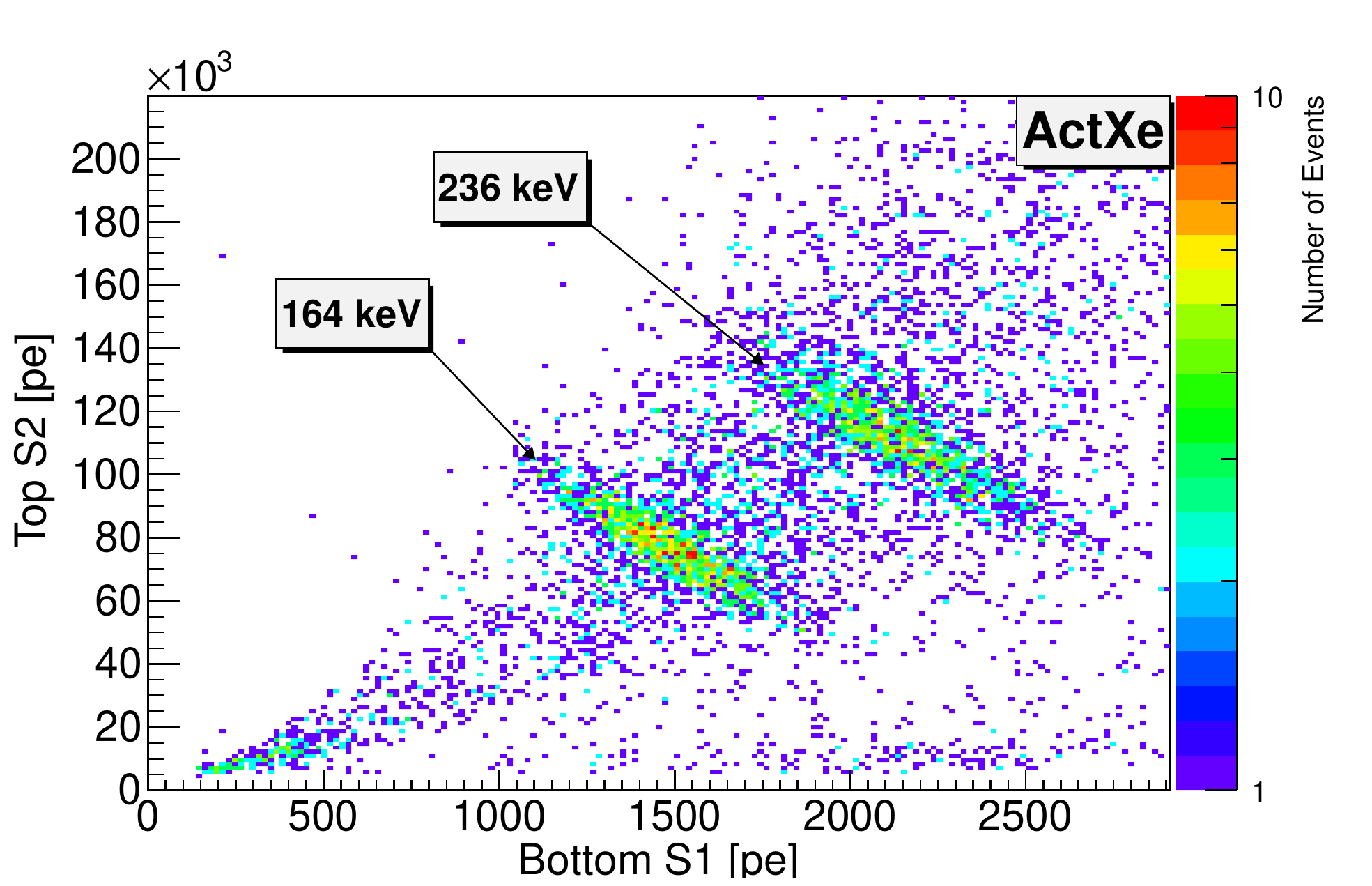}\\
\includegraphics[width=0.47\textwidth]{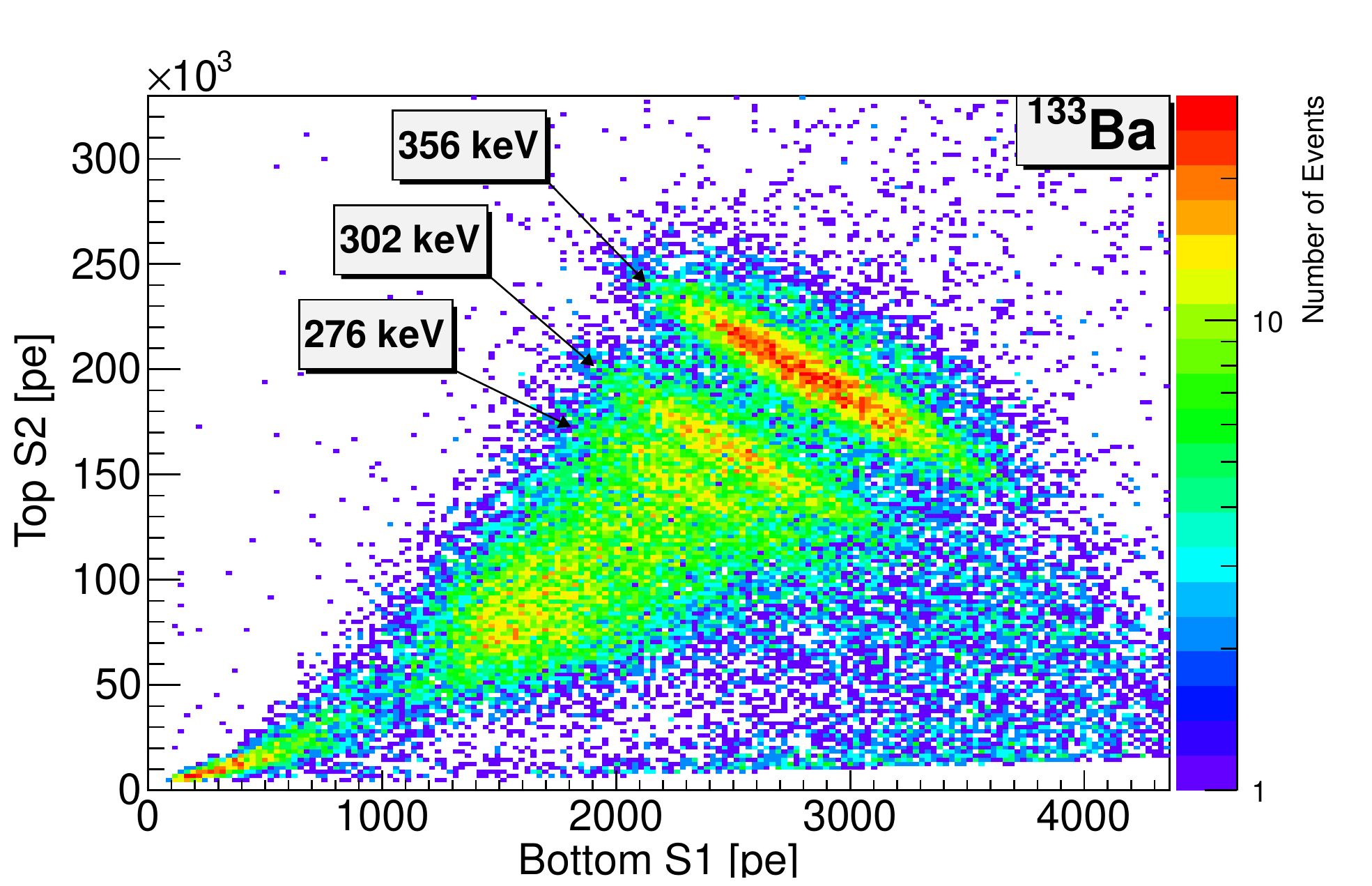} \hspace*{3mm}
\includegraphics[width=0.47\textwidth]{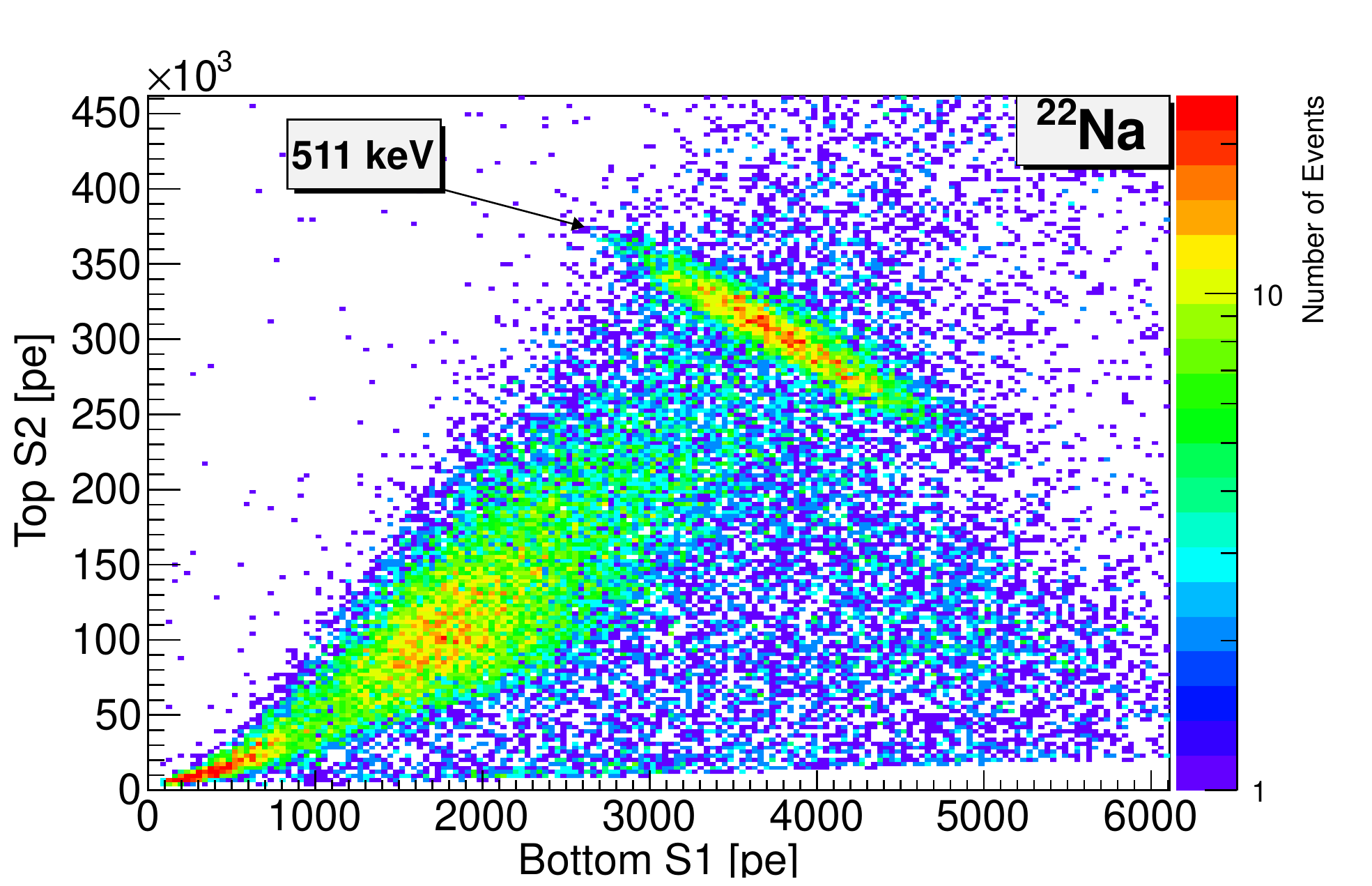} \\
\includegraphics[width=0.47\textwidth]{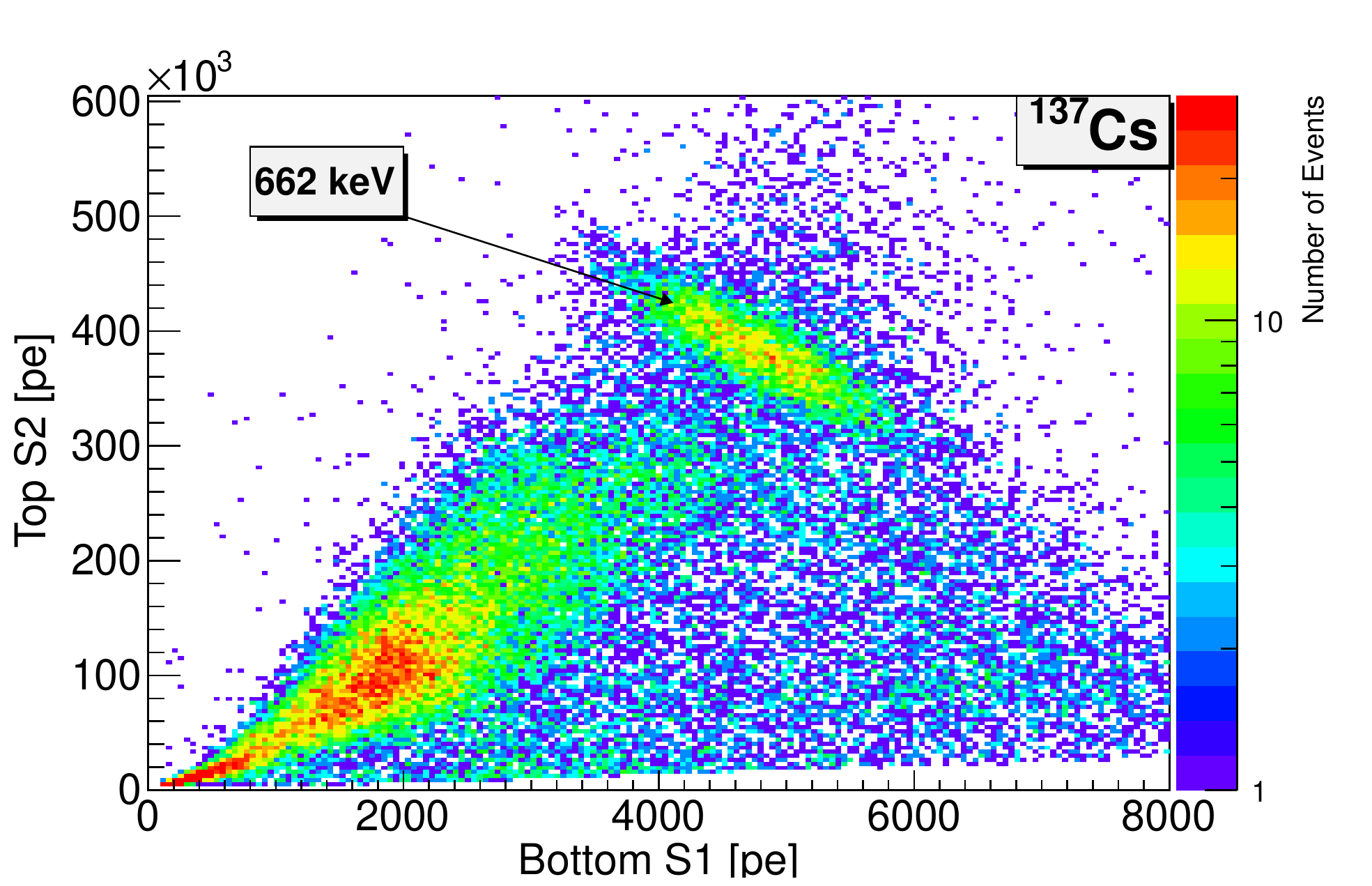} \hspace*{3mm}
\includegraphics[width=0.47\textwidth]{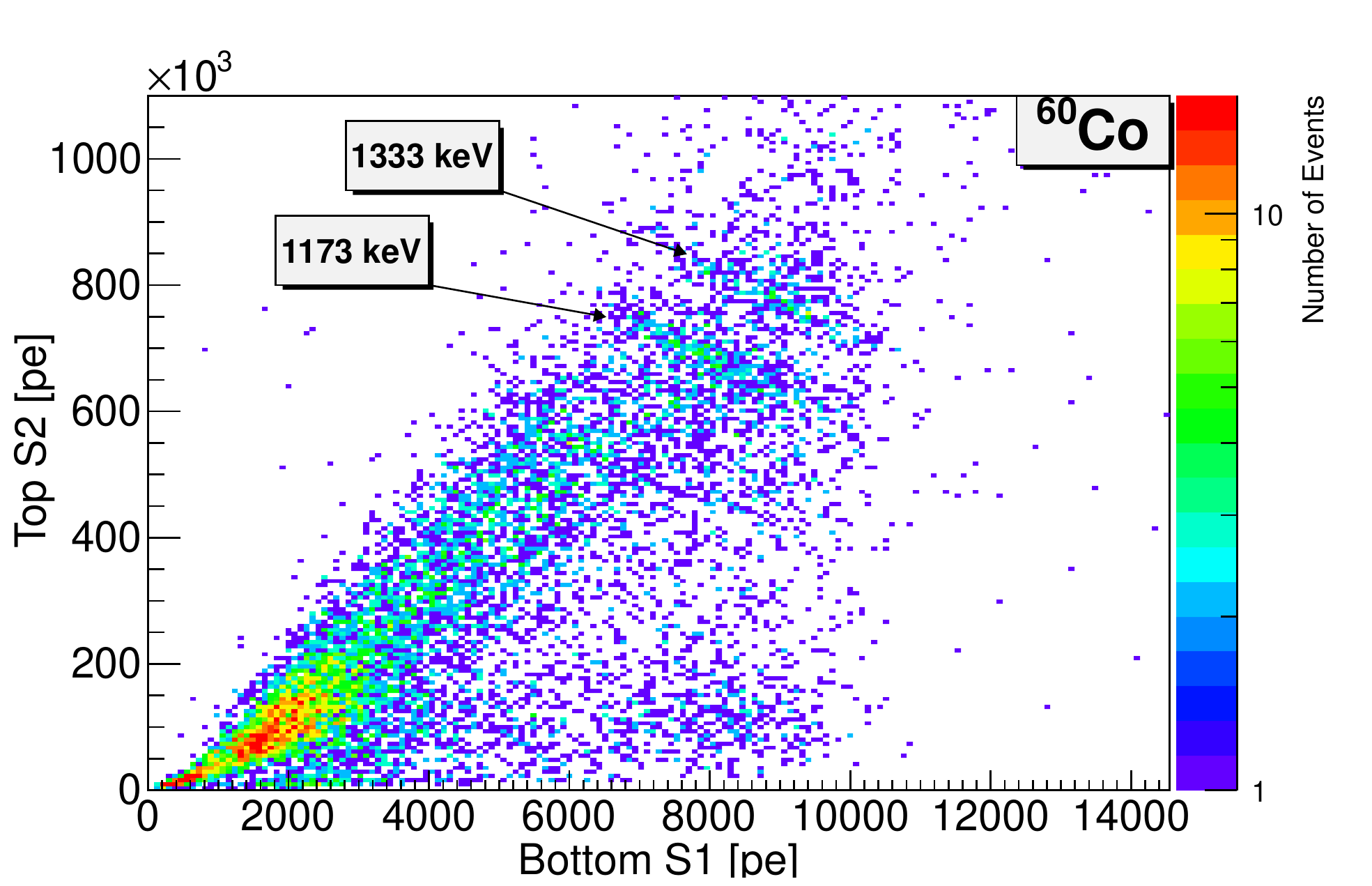}
\caption{Correlation of the Top S2 and Bottom S1 signals for all the sources used in this study taken at a drift field of 200\,V/cm. This set of data was used to determine the $\alpha$ and $\beta$ values. From top left to bottom right the plots represent spectra for $^{57}$Co, $^{131m}$Xe and $^{129m}$Xe, $^{133}$Ba, $^{22}$Na, $^{137}$Cs, and $^{60}$Co. Note that the TS2 and BS1 ranges increase with energy for each plot but the ratio of TS2 to BS1 ranges are held constant for all sources.}
\label{fig:all-sources}
\end{figure}
\begin{figure}[tbp]
\centering
\includegraphics[width=.6\textwidth]{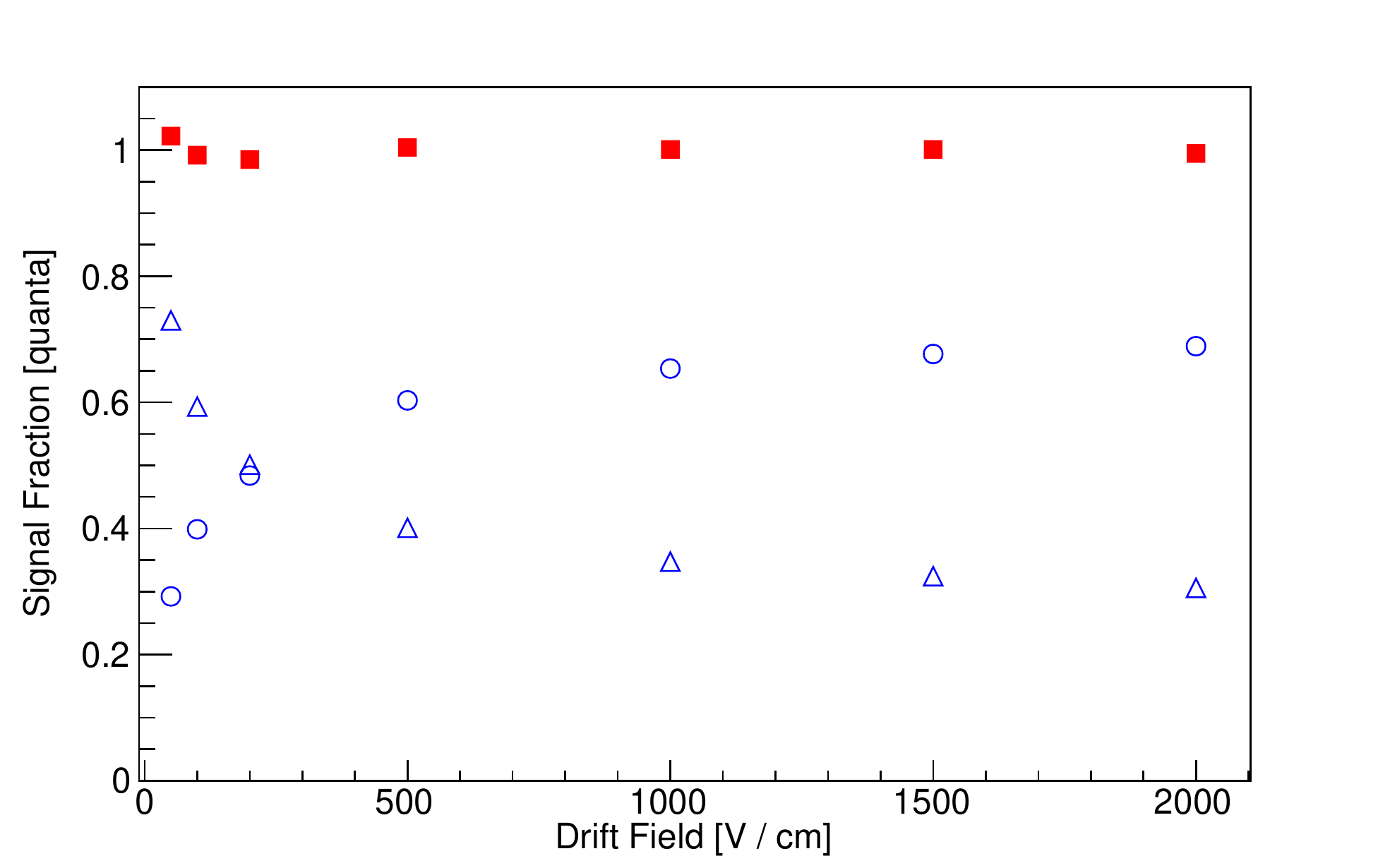}
\caption{Field dependence of scintillation (BS1 as open triangles) and ionization (TS2 as open circles) yield and their sum (filled squares) in LXe for 356\,keV electron recoils from $^{133}$Ba. The values are normalized to unity at 2,000\,V/cm.}
\label{fig:S1-S2-field-dependence}
\end{figure}

Since there is at least one photo-absorption peak associated with each radioactive source, the response of the MiX detector to mono-energetic events can be measured for each source.  Figure~\ref{fig:all-sources} shows a 2D representation of the TS2 versus BS1 signals recorded for each source at a drift field of 200\,V/cm. The intrinsic event-by-event anti-correlation of light (BS1) and charge (TS2) signals in LXe is clearly seen for each source.  There are two manifestations (macroscopic and microscopic) of anti-correlations to be considered. The mean macroscopic anti-correlation~\cite{Aprile-07} in BS1 and TS2 is seen when sweeping the drift field for a specific gamma line energy (as shown in Fig.~\ref{fig:S1-S2-field-dependence}). As the drift field is increased, ionization collection grows and eventually asymptotes while light signal drops complementarily. The behavior is due to ionization electrons being collected rather than forming additional S1 recombination signal. The microscopic anti-correlation, due to the event-by-event stochastic nature of electron recoil energy partitioning into excitation and ionization (with the effects of recombination), coupled with the assumed constant work function of LXe ($W=13.7$\,eV)~\cite{Szydagis-11,Szydagis-13,Dahl-PhD} and high detector light yield, can be seen for the various data sets in Fig.~\ref{fig:all-sources}. The constant nature of $W$ (the average energy to produce a scintillation photon or to liberate an electron) across energy is approximately verified, but precision measurements are necessary to confirm this behavior.

\begin{figure}[tbp]
\centering
\includegraphics[width=0.47\textwidth]{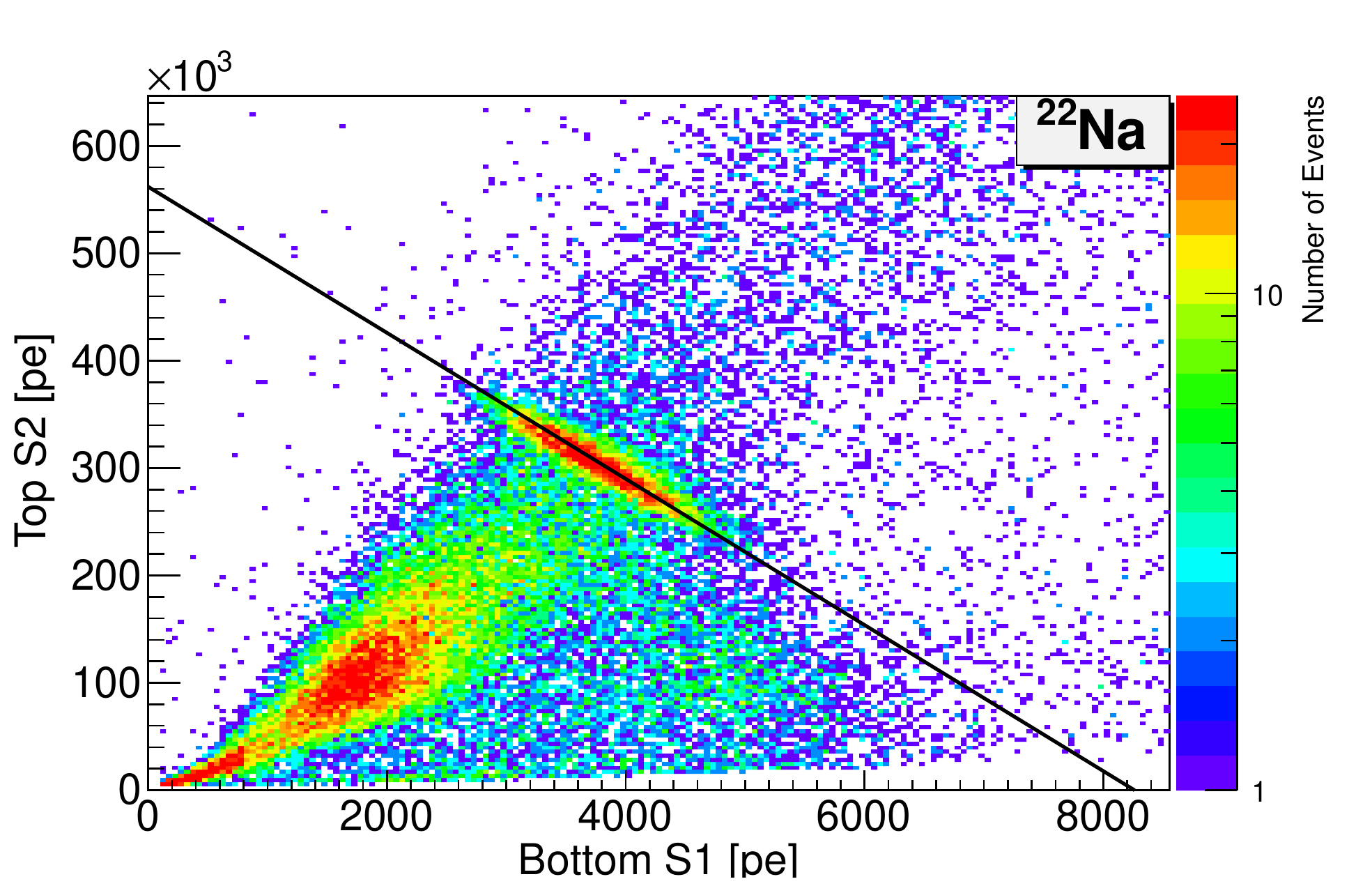} \hspace*{3mm}
\includegraphics[width=0.47\textwidth]{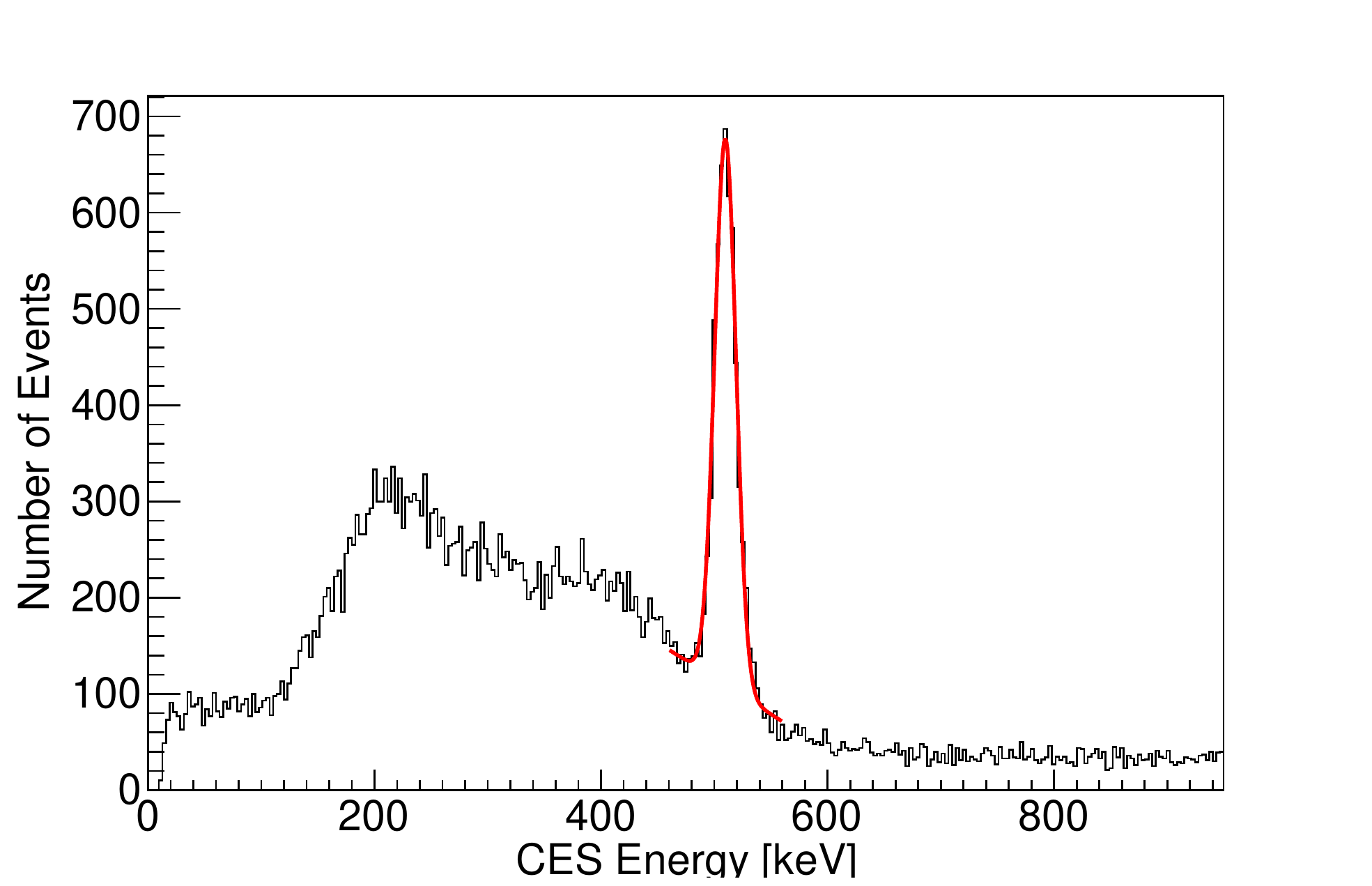}
\caption{Energy resolution for large fiducial volume selection as shown in Fig.~5, but no $XYZ$ position corrections applied. Left panel: Correlation of the Top S2 and Bottom S1 signals for $^{22}$Na in a 200\,V/cm drift field. A 2D Gaussian fit was applied to the S2-S1 profiles to find the anti-correlation axis (black diagonal line). Right panel: The combined energy spectrum showing the 511\,keV line together with Gaussian fits plus linear background are shown to give an energy resolution ($\sigma$/E) of 1.84\%. With $XYZ$ position corrections applied the energy resolution is 1.62\%.}
\label{fig:projection-wide}
\end{figure}

Although both the BS1 and TS2 signals can be used individually to determine the energy of an interaction, it is advantageous to combine the BS1 and TS2 signals for energy reconstruction to achieve an optimized energy resolution. The combined energy scale $E_{CES}$ is determined by combining the two signals to form
\begin{equation}
\label{eq:E_CES}
E_{CES}=W\left(\frac{BS1}{\alpha}+\frac{TS2}{\beta}\right),
\end{equation}
where $\alpha$ (also referred to as $g_{1}$, PDE) is the average photon detection efficiency and $\beta$ (or $g_{2}$, ${ETE}\cdot{EEE}\cdot{SE}$) is the product of the electron extraction efficiency (EEE), the electron transmission efficiency while passing by the gate grid (ETE), and the single electron S2 gain (SE). This combination of signals provides the optimal energy resolution at the energies studied in this work. Figure~\ref{fig:projection-wide} (right panel) shows an example of a combined energy spectrum for $^{22}$Na in a 200\,V/cm drift field. It corresponds to the projection of events in the TS2 vs. BS1 spectrum (Fig.~\ref{fig:projection-wide} left panel) along the S2-S1 anti-correlation axis whose slope is given by the ratio of $-\beta/\alpha$. Using the large $XY$ fiducial region and a $9\,\mu s>Z>4\,\mu$s drift time cut, an energy resolution ($\sigma$/E) of 1.62\% for the 511\,keV gamma energy line is obtained. It is worth noting that when applying no $XYZ$ position corrections, the energy resolution of the 511\,keV line is still 1.84\%.

In order to determine the $\alpha$ and $\beta$ values (see Eq.~\ref{eq:E_CES}), 2D Gaussian fits were applied to the S2-S1 profiles of each of the six spectra displayed in Fig.~\ref{fig:all-sources} to find their anti-correlation axes. Using the BS1 and TS2 intercepts, BS1$_I$ and TS2$_I$, of the anti-correlation axis in each spectrum,  $\alpha$ and $\beta$ values were determined for each photo-absorption peak according to $\alpha = W ( \text{BS1} _{I} / E )$ and $\beta = W ( \text{TS2} _{I} / E )$, where $E$ is the energy of the photo-absorption events. Averaging over all sources yields $\alpha=(0.239\pm0.012)$\,pe/photon and $\beta=(16.1\pm0.6)$\,pe$/e^{-}$ for the MiX detector at a drift field of 200\,V/cm.

The zero field light yield of the detector, when extrapolated from the anti-correlation axis to the BS1 intercept, is $(17.1\pm 0.7)$\,pe/keV at $122$\,keV. A correction factor of 0.89~\cite{Dahl-PhD} is applied to the extrapolated light yield to obtain an estimated true zero field light yield of $(15.2\pm0.6)$\,pe/keV at $122$\,keV. This corresponds to 64\,photons/keV, in agreement with NEST~\cite{Szydagis-11}.

\begin{figure}[tbp]
\centering
\includegraphics[width=0.47\textwidth]{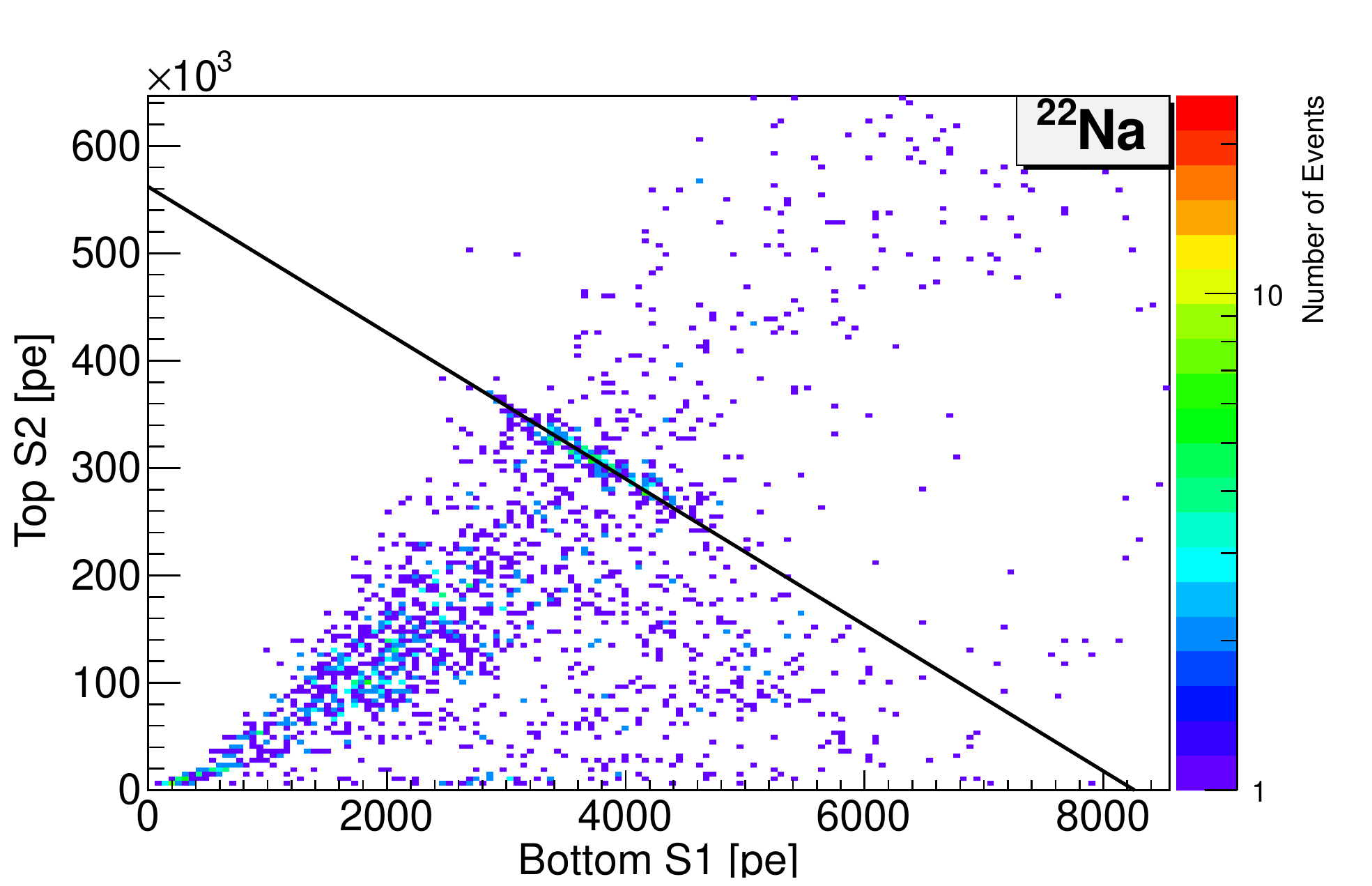} \hspace*{3mm}
\includegraphics[width=0.47\textwidth]{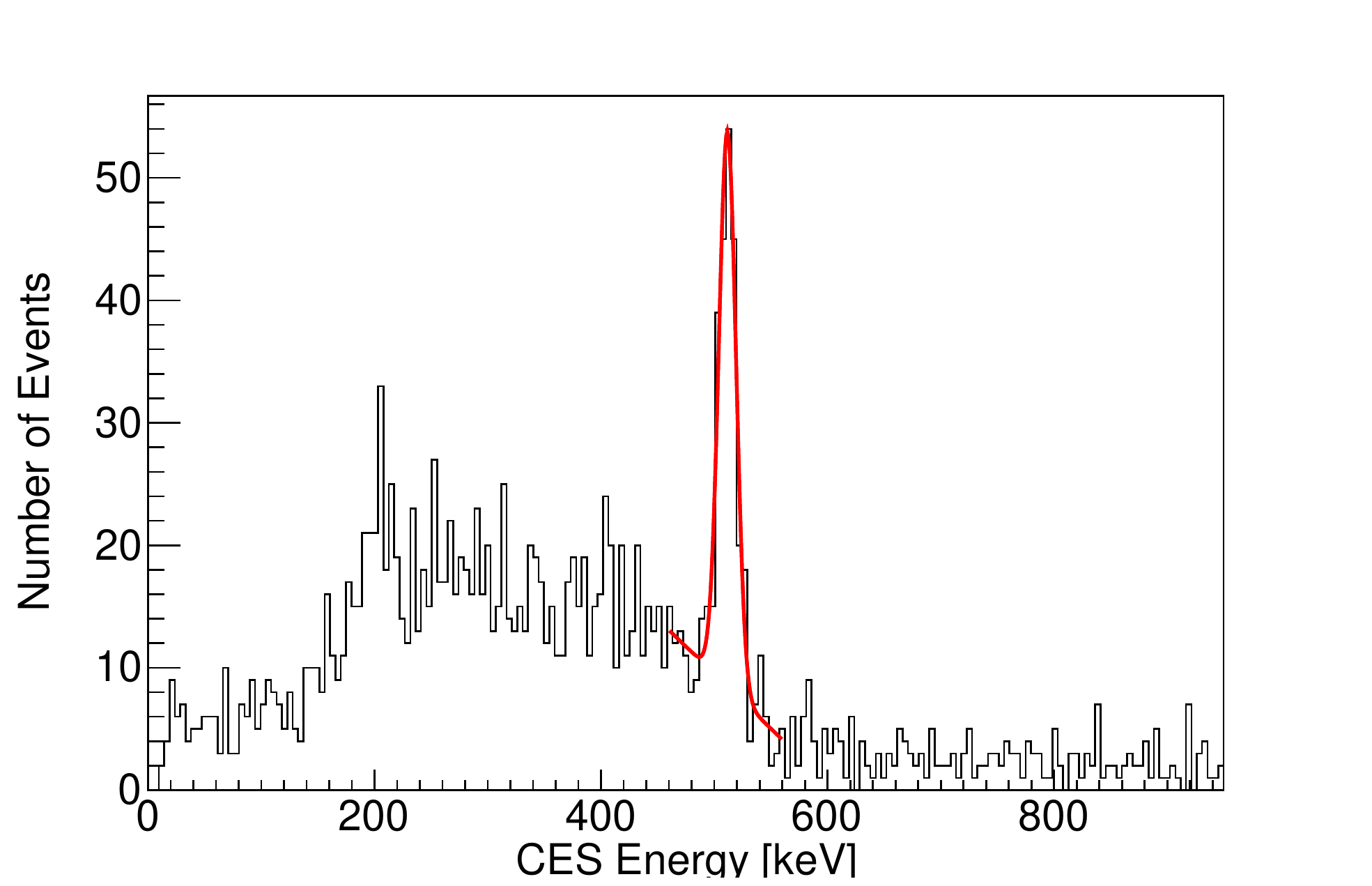}
\caption{Energy resolution for small fiducial volume selection, with $XYZ$ position corrections applied, otherwise like Fig.~9. The ultimate energy resolution obtained for the 511\,keV line is 1.36\%.}
\label{fig:projection-narrow}
\end{figure}

The result of high light yield is good energy resolution and discrimination power between nuclear and electron recoils over a large energy range. Energy resolution as a function of energy is investigated in this work by making fiducial $XY$ cuts (see Fig.~\ref{fig:xy-pos}) to select events from the central part of the detector for each source. The resulting energy spectra for each source were fit with background plus gaussian functions. Figure~\ref{fig:projection-narrow} shows the results from the annular-shaped fiducial $XY$ region (bounded by the two red circles) yielding a resolution of $\sigma/\rm{E} = (1.36\%\pm0.07)\%$. This region corresponds to a fiducial volume of 1.1\,gram of LXe, a small fraction of the entire sensitive volume of 114\,grams. The restricted annular region is selected for its superior energy resolution while still providing enough photo-absorption events to allow a satisfactory signal fit. The improvement in energy resolution can be attributed to the axial symmetry of this centralized region overcoming the effects of a statistics-limited correction map.

\begin{figure}[tbp]
\centering
\includegraphics[width=.65\textwidth]{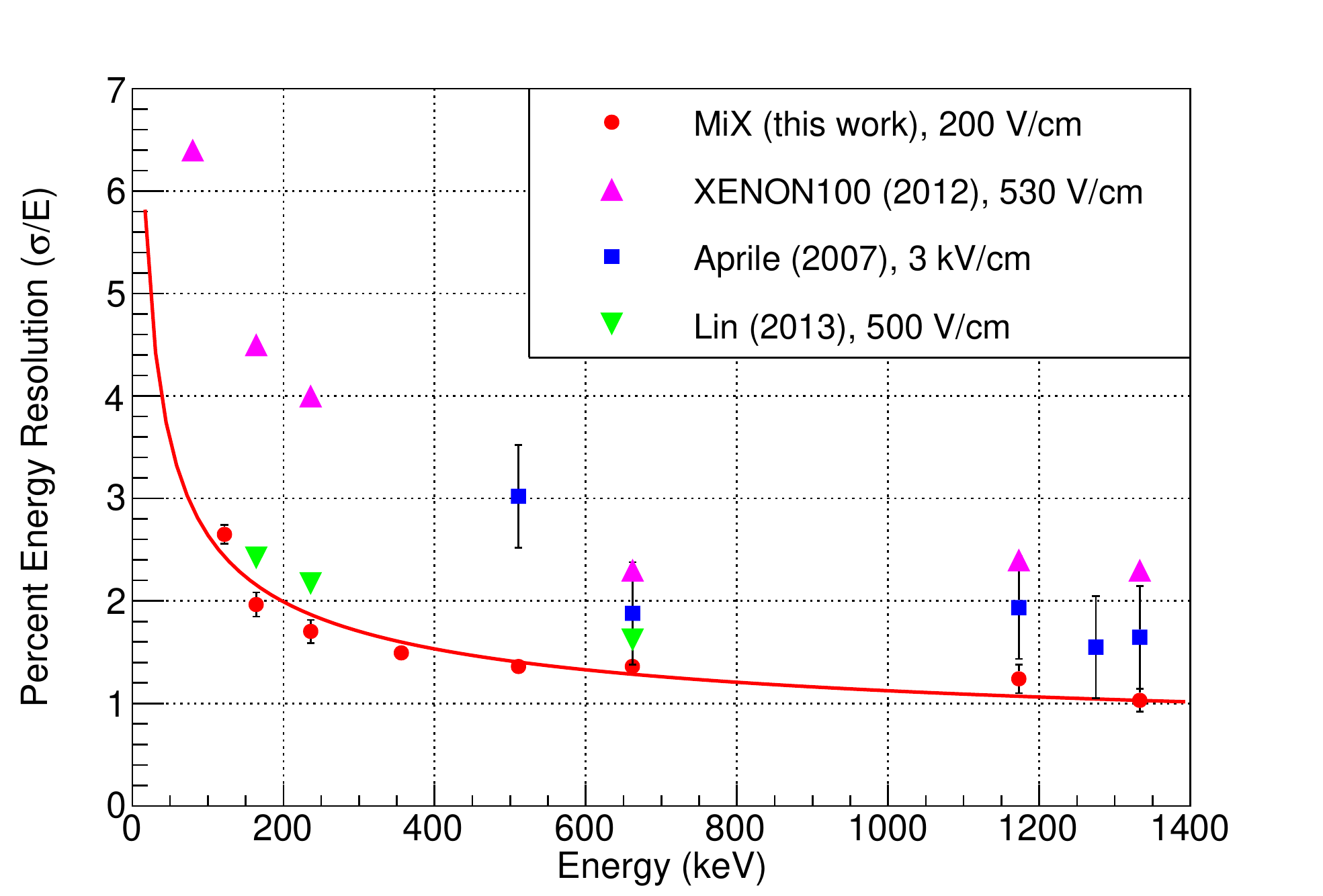}
\caption{Energy resolution in liquid xenon measured by Aprile et al.~\cite{Aprile-07}, XENON100~\cite{XENON100-12}, Lin et al.~\cite{Lin-14} and this work. The solid (red) line represents the fit result to the data above 100\,keV to a function $\sigma$/E$ = (22.2/\sqrt{E} + 0.42)\%$.}
\label{fig:E-resolution}
\end{figure}

The energy resolution results for all sources are shown in Fig.~\ref{fig:E-resolution} with an ultimate resolution of $\sigma/\rm{E} =(1.03\pm0.11)\%$ at 1.33\,MeV, which may be the best achieved in liquid xenon to date. Also shown in Fig.~\ref{fig:E-resolution} are results obtained in other LXe detectors around the world~\cite{Aprile-07,XENON100-12,Lin-14}. The performance of XENON100~\cite{XENON100-12} is typical of early large scale dual-phase instruments, which did not have high photon detection efficiency, while the smaller scale R\&D devices represent what can be achieved with advancement in TPC design, operation, and sensor performance. The early data by Aprile~\cite{Aprile-07} were collected in a single phase detector using charge amplifiers instead of PMTs for the proportional signal. Differences in the resolution results of small-scale R\&D devices can be attributed to varied quanta collection efficiency from differences in electrode optical transparency, TPC aspect ratio, and PMT performance. The excellent energy resolution obtained in the MiX detector can be attributed to the very high light collection efficiency, stability in the electric fields, PMT gains, and LXe level as well as the ability to reconstruct $XY$ position with high precision, allowing for the selection of a fiducial volume that avoids saturation and degeneracies typically observed in small LXeTPCs with high symmetry.

\begin{table}[tbp]
\caption{Selection of radioactive sources together with respective energy resolutions are shown.}
\label{tab:radioactive-sources}
\smallskip
\centering
\begin{tabular}{|ccc|}
\hline
Calibration Source & Energy & Resolution  \\
                   & [keV]  & (\%) \\
\hline
\hline
$^{57}$Co   & 122   & $2.65\pm0.09 $ \\
$^{131m}$Xe & 164   & $1.97\pm0.12 $ \\
$^{129m}$Xe & 236   & $1.70\pm0.11 $ \\
$^{133}$Ba  & 356   & $1.49\pm0.07 $ \\
$^{22}$Na   & 511   & $1.36\pm0.07 $ \\
$^{137}$Cs  & 662   & $1.36\pm0.06 $ \\
$^{60}$Co   & 1,173  & $1.24\pm0.14 $ \\
$^{60}$Co   & 1,333  & $1.03\pm0.11 $ \\
\hline
\end{tabular}
\end{table}

\section{Conclusions}

We report the design and performance of the Michigan Xenon detector, a small 3D position sensitive dual-phase liquid xenon time projection chamber with high light yield, long electron lifetime, exquisite energy resolution over a wide dynamic range and stable operation with minimal human interaction. The ultimate energy resolution of approximately 1\% ($\sigma/$E) at 1.33\,MeV may be the best resolution achieved in liquid xenon over a wide energy range to date.  If similar energy resolution is desired in a large detector, efforts must be made to improve upon photon detection efficiency and increase sensor coverage while maintaining current large scale charge detection capabilities.

As the MiX detector is refined, much work on further characterizing LXe as a high
fidelity radiation imaging medium is on the horizon. A future important goal is to precisely quantify the ER/NR discrimination capabilities of a LXeTPC under varied conditions. There are many free parameters to optimize in a LXe detector, such as internal geometry,  pressure, drift and extraction fields, and LXe level. And there is still a question of how these parameters play a role in determining the ER/NR discriminating power in LXe signal production.%, which is the degree of confidence with which one can categorize an event in LXeTPC as either NR or ER.

Ultimately, the modern scientific goals of LXeTPCs are to use the technology to discover new physics such as WIMP dark matter and neutrinoless double beta decay. Physicists may be closer than ever to achieving these feats. Additionally, due to the incredible resolving power of these detectors, both in energy and in position, additional important applications in LXe-assisted tomography and national security are within reach.

\acknowledgments

We thank Karl Giboni for the use of the cryostat design and gas purification layout along with guidance in system operation. We also would like to acknowledge Jianglai Liu for supplying PMTs and the BPMT base, and Fei Gao, Mengjiao Xiao, and Daniel Morton for useful discussions. K. Ni and Q. Lin acknowledge support from the National Natural Science Foundation of China No.~11375114. This work was largely supported by generous funds from the University of Michigan.

\appendix
\section{Reconstruction Singularities in Multi-PMT detectors}
\label{sec:theorem}

It was discovered during analysis of data from a Monte Carlo simulation of the MiX detector that a degeneracy exists in the position reconstruction of the detector. Eventually this was formulated as a general topological theorem (see Theorem~\ref{thm:degen}). This degeneracy restricts the ability to reconstruct event positions inside a detector in certain regions of space, which in turn may limit the ability to make position-dependent signal corrections and fiducial volume selections. Introducing a large enough asymmetry into the PMT arrangement can ensure that the degeneracy does not occur.

\subsection{Position-Signal Degeneracy Theorem}

\begin{figure}[tbp]
\centering
\includegraphics[width=.5\textwidth]{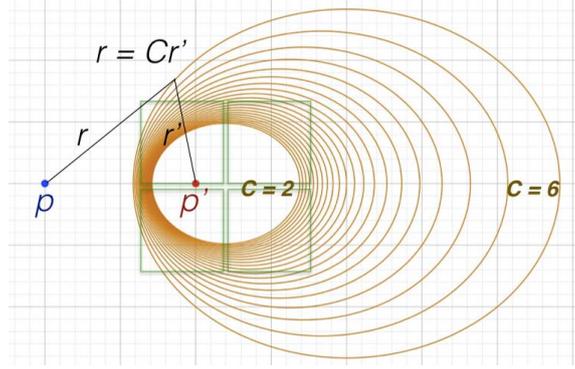}
\caption{Application of the Position-Signal Degeneracy Theorem to the top PMT geometry of the MiX detector. Two events occur at points $p$ (in blue) and $p'$ (in red). Loci of points satisfying the equation $r = Cr'$ are shown in orange for values of $C$ between 2 and 6. Since the MiX TPMTs (green squares) fall over several of these loci, there will be many pairs of indistinguishable points $p$, $p'$.}
\label{fig:theorem-1}
\end{figure}

For a given event, the number of light quanta that strike each PMT is referred to as the hit pattern. Since the hit pattern is determined by the position and energy of the event itself, it is natural to ask whether the position and energy of the event are uniquely determined by the hit pattern. In general, this is not the case. Even with an arbitrary number of PMTs, it is possible for two events at distinct positions to produce indistinguishable \emph{normalized} hit patterns. If normalization is necessary for the position reconstruction model, then theorem~\ref{thm:degen} applies. A visual representation of how theorem~\ref{thm:degen} limits the position reconstruction capabilities of a detector with MiX-like geometry is illustrated schematically in Fig.~\ref{fig:theorem-1}.

\begin{figure}[tbp]
\centering
\includegraphics[width=.6\textwidth]{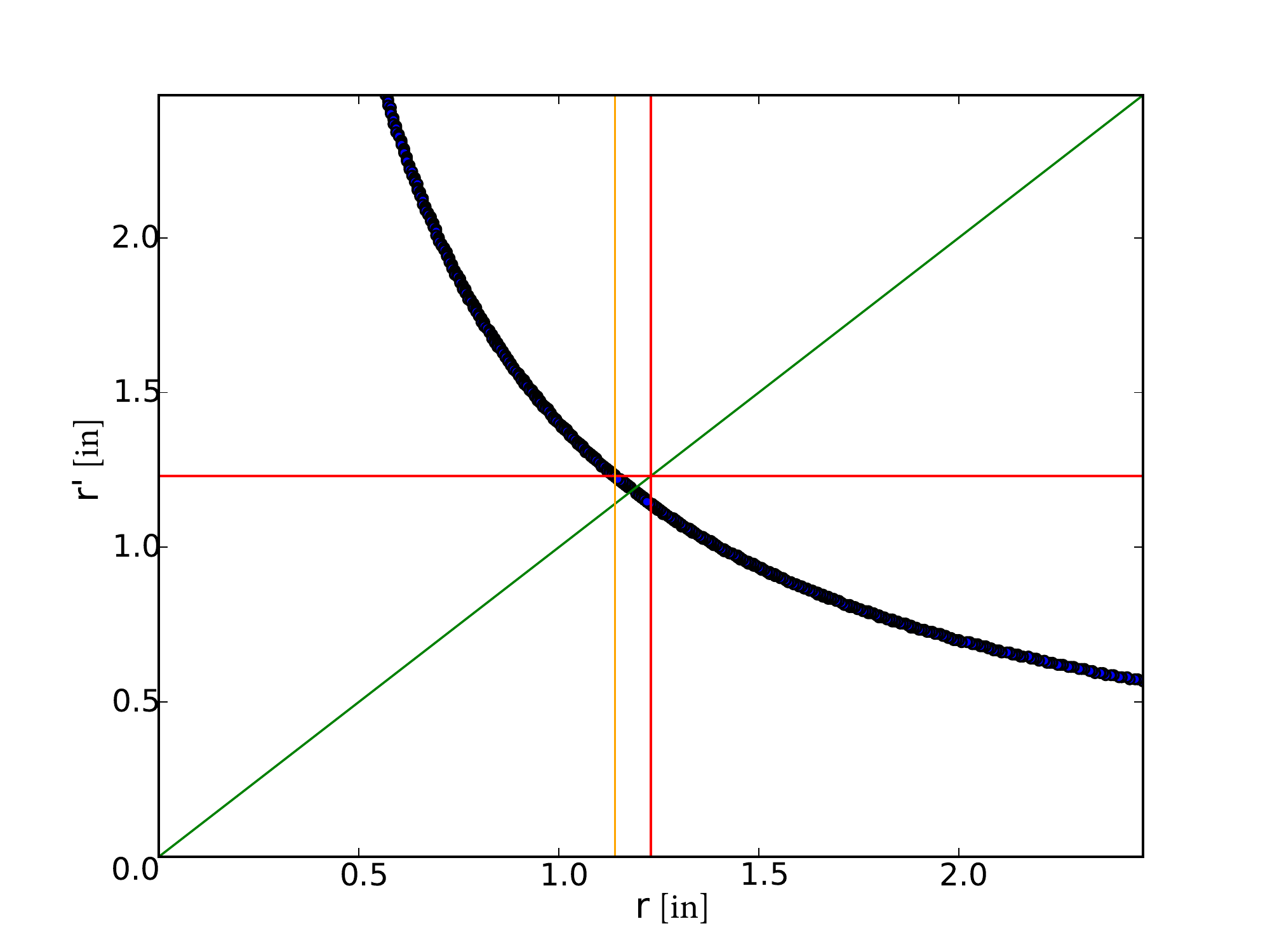}
\caption{Numeric degenerate pairing results from simulated four PMT array. The black curve shows pairs of radii $r$, $r'$ in inches from the center of the TPC that are degenerate. Red lines demarcate the maximum radius of a MiX S2 event. The green line is $r=r'$ to provide scale. The orange line marks the lowest radius for which the degenerate pair radius lies within the MiX S2 region, at 29.0\,mm.}
\label{fig:theorem-2}
\end{figure}

In the MiX detector geometry, it can be shown that for every S2 event, there is a twin S2 event that would produce an indistinguishable hit pattern from a different event position. However, since the volume in which S2 events may occur is finite, there is a subset of points inside the detector whose degenerate partners can never occur. Using numerical methods, one may locate these degenerate event positions and place bounds on the spatial region of the detector with good position reconstruction. For the MiX detector, the maximum radius with no degeneracy is 29.0\,mm (as shown in Fig.~\ref{fig:theorem-2}). This radius is large enough to avoid position degeneracy in the fiducial volume used for this analysis (see Fig.~\ref{fig:xy-pos}).  %However, this maximum radius can not be realized as a fiducial cut due to the associated heavy lensing effects (see Fig.~5) coupled with position uncertainties at high radius.%rMaxNoDegen =  1.14038012671

\begin{theorem}
Let an event occur at some point $p \in \mathbb{R}^3$ with magnitude
$M$, producing a signal which propagates uniformly through space,
so that a measuring device with distance $r$ to $p$ will register
a response of magnitude $M / r^2$. Suppose that the signal magnitude
due to this event is measured at some points $\{ m_i \}$ of distances
$r_i$ to $p$.

If there exists a point $p'$ whose distances $r_i'$ to each of the
measurement points $m_i$ satisfy $r_i' = C r_i$ for some constant
$C$, then it will be impossible to distinguish an event at point
$p$ from an event at point $p'$ if the absolute magnitude of the signal is unknown.
\label{thm:degen}
\end{theorem}

\noindent\textbf{Proof: }\\
Given the conditions in theorem \ref{thm:degen}, we may choose magnitude
$M' = M \cdot C^2$ such that $\forall i$,
\begin{equation}
\frac{M'}{r_i'^2} = \frac{M \cdot C^2}{r_i'^2} = \frac{M}{r_i^2}.
\end{equation}
An event of magnitude $M'$ at the point $p'$ will produce exactly
the same signal at each measuring point, making it indistinguishable
from the event at $p$. The finite optical position resolution of detectors only increases the number of available radii which satisfy this condition.

\end{document}